\documentclass[iop]{emulateapj}
\usepackage{amsmath}
\usepackage{graphicx}
\usepackage{natbib}

\shorttitle{Multiple populations in Fornax GCs}
\shortauthors{S. S. Larsen et al.}
\slugcomment{ApJ, accepted (27 Aug 2014)}

\begin{document}
\title{Nitrogen abundances and multiple stellar populations in the globular clusters of the Fornax dSph}

\author{S{\o}ren S. Larsen}
\affil{Department of Astrophysics/IMAPP, Radboud University Nijmegen, PO Box 9010, 6500 GL Nijmegen, The Netherlands}
\email{s.larsen@astro.ru.nl}

\author{Jean P. Brodie}
\affil{UCO/Lick Observatory, University of California, Santa Cruz, CA 95064, USA}

\author{Frank Grundahl}
\affil{Stellar Astrophysics Centre, Department of Physics and Astronomy, Aarhus University, Ny Munkegade 120, DK-8000 Aarhus C, Denmark}

\and

\author{Jay Strader}
\affil{Department of Physics and Astronomy, Michigan State University, East Lansing, Michigan 48824, USA}

\begin{abstract}
We use measurements of nitrogen abundances in red giants to search for multiple stellar populations in the four most metal-poor globular clusters (GCs) in the Fornax dwarf spheroidal galaxy (Fornax~1, 2, 3, and 5). New imaging in the F343N filter, obtained with the Wide Field Camera 3 on the Hubble Space Telescope, is combined with archival F555W and F814W observations to determine the strength of the NH band near 3370~\AA . After accounting for observational errors,
the spread in the F343N-F555W colors of red giants in the Fornax GCs is similar to that in M15 and corresponds to an abundance range of $\Delta\mathrm{[N/Fe]}\sim2$ dex, as observed also in several Galactic GCs. The spread in F555W-F814W is, instead, fully accounted for by observational errors.
The stars with the reddest F343N-F555W colors (indicative of N-enhanced composition) have more centrally concentrated radial distributions in all four clusters, although the difference is not highly statistically significant within any individual cluster. From double-Gaussian fits to the color distributions we find roughly equal numbers of  ``N-normal'' and ``N-enhanced'' stars (formally $\sim40$\% N-normal stars in Fornax~1, 3, and 5 and $\sim60$\% in Fornax~2).
We conclude that GC formation, in particular regarding the processes responsible for the origin of multiple stellar populations, appears to have operated similarly in the Milky Way and in the Fornax dSph. Combined with the high ratio of metal-poor GCs to field stars in the Fornax dSph, this places an important constraint on scenarios for the origin of multiple stellar populations in GCs.
\end{abstract}


\section{Introduction}

It was noted decades ago that a large fraction of stars in globular clusters (GCs) have anomalous abundances of several light elements compared to field stars \citep{Cohen1978,Kraft1979,Norris1981,Gratton2004}. While the anomalous abundance patterns were first found in the brightest giants that may have undergone internal mixing \citep[e.g.][]{Sneden1986,Langer1993}, stellar evolutionary effects have since been ruled out as the main underlying cause because unevolved stars display many of the same anomalies \citep{Briley1994,Briley1997,Grundahl2002,Gratton2004,Cohen2005}. Bimodal distributions of CN line strengths have been observed for sub-giants and stars at the turn-off \citep{Kayser2008}, while high-precision photometry has revealed multiple main sequences, split sub-giant and giant branches, and other features in GC color-magnitude diagrams that indicate complex formation and/or internal chemical enrichment histories \citep{Gratton2012}.
 
One of the most common and best-known anomalies is the anti-correlation between the abundances of Na and O. This phenomenon is so common that it has even been suggested that GCs could be \emph{defined} as clusters that display the Na-O anticorrelation \citep{Carretta2010}. Other observed patterns characteristic of GCs include anti-correlated C-N and O-N abundances, as well as a N-Na correlation
\citep{Cottrell1981,Sneden2004,Yong2008,Carretta2009a}. The relations involving N are of great practical importance because N abundance variations can be detected photometrically with relative ease due to the strong NH absorption bands in the ultraviolet \citep{Grundahl1998,Grundahl2002}.

The evidence outlined above is often interpreted as pointing towards a scenario in which GCs consist of an initial population of stars that formed by normal processes (with composition similar to that observed in field stars) and a second population ``polluted'' by some mechanism specific to GCs. An important additional constraint comes from observations of heavier elements, such as Ca and Fe, that are produced in supernovae. Within the Milky Way GC system, large internal variations in abundances of these elements have only been observed in a small, albeit growing, number of clusters \citep[$\omega$ Cen, M2, M22 and M54;][]{DaCosta2009,Johnson2009,Carretta2010a,Yong2014a}, with a few known cases also in M31 \citep{Fuentes-Carrera2008}. While there is increasing evidence that measurable spreads in [Fe/H] may be more common in GCs than previously assumed \citep{Carretta2009c,Willman2012}, the spread remains small or undetectable within tight observational limits ($\Delta\mbox{[Fe/H]}\la0.05$ dex) in most clusters \citep{Yong2008b,DErcole2010,Cohen2011,Carretta2014,Yong2014}. Massive asymptotic giant branch (AGB) stars or massive main sequence stars (single or binary), in which light-element abundances can be modified by proton-capture nucleosynthesis at high temperatures, thus appear to be the most plausible main sources of polluted gas \citep{Decressin2007,DErcole2008,Renzini2008,deMink2009}, with supernovae playing a negligible or minor role for self-enrichment in most clusters. Self-enrichment by SNe may have been important in the few GCs that show significant spreads in $\mathrm{[Fe/H]}$ and it may also be responsible for the color-luminosity relation of metal-poor GCs, also known as the ``blue tilt'', that is observed in some extragalactic GC systems \citep{Harris2006a,Mieske2006,Strader2006}.

One of the main challenges is to explain the very large fractions of stars with anomalous composition that are typically observed. Indeed, the ``normal'' stars are often the \emph{minority} \citep[e.g.,][]{DAntona2008,Milone2012a}, which makes it difficult to understand how a second generation could have formed out of ejecta from first-generation stars. For a normal stellar initial mass function (IMF), the ejecta produced by massive AGB stars account for only $\sim5$ percent of the total initial mass of a population \citep{DErcole2008}. Even if this ejected matter were able to form stars at 100\% efficiency, the second generation  should, therefore, at most constitute a few percent of the total mass. A commonly proposed solution to this \textit{mass-budget} problem is that the first generation of stars was initially much more populous than it is now.
To explain the observed large fractions of second-generation stars, most scenarios require that GCs were at least a factor of 10 more massive initially than is currently observed. In this picture, present-day GCs are merely the remnants of initially far more massive systems, which have now \emph{preferentially} lost most of their first-generation stars \citep{Vesperini2010,Bekki2011,Valcarce2011,Ventura2013}. In contrast, it is interesting to note that the SN self-enrichment scenarios that aim to explain the blue tilt do not suffer from a mass-budget problem \citep{Strader2008,Bailin2009,Goudfrooij2014}.

Currently, about 1\%--2\% of the stars in the Galactic halo belong to GCs, but if GCs did indeed lose the vast majority of their first-generation stars, then up to $\sim50\%$ of the halo stars may have formed within GCs \citep{Martell2010,Martell2011}. This estimate takes into account that second-generation stars have also been lost from GCs due to evaporation or complete dissolution, thus accounting for the $\sim3\%$ of CN-strong stars observed in the halo. While this already implies a very high GC formation efficiency for the halo, \emph{dwarf galaxies} can have GC specific frequencies far exceeding that of the Milky Way \citep{Miller2007,Peng2008,Georgiev2010,Harris2013}.
This turns out to provide a strong constraint on GC formation scenarios, because the present-day ratios of field stars to GCs observed in some dwarfs appear too low to accommodate an amount of mass loss comparable to that implied for Galactic GCs. Specifically, based on a detailed comparison of GC and field star metallicity distributions, we have found that GCs currently account for about 1/5--1/4 of the metal-poor stars in the Fornax dwarf spheroidal galaxy  \citep{Larsen2012}. 
The GCs in Fornax could therefore not have been more than a factor of $4-5$ more massive initially (for a standard IMF), or the lost metal-poor stars should now be present in the field. These stars are not observed. A similar low field/GC ratio is found in the Wolf-Lundmark-Melotte galaxy, and the 5 GCs in  the IKN dSph  may even account for half or more of  the metal-poor stars in that galaxy \citep{Larsen2014}. 
The constraints on GC mass loss from these observations would clearly be less strong if a significant fraction of the metal-poor field stars have subsequently been lost from the dwarfs, but at least in the cases of Fornax and WLM, this appears unlikely as these galaxies are both quite isolated \citep{Sandage1985,Minniti1996,Penarrubia2009}.

Recently, GC formation scenarios have been put forward that might suffer from a less severe mass budget problem. 
\citet{DAntona2013} suggested that it may be possible to (barely) accommodate the low field/GC ratio in Fornax within the AGB scenario if the second-generation IMF is bottom-heavy so that essentially all second-generation stars formed are still alive today. However, this still leaves little room for the formation of any additional stars at the same metallicities as the GCs, either in the form of (now disrupted) low-mass clusters or bona-fide field stars. \citet{Bastian2013a} have proposed that the abundance anomalies observed in GCs may result from the accretion of ejecta from massive interacting binary stars onto proto-stellar discs of low-mass stars during the first 5--10 Myr of the cluster evolution. The mass-budget problem  is much less severe is this scenario, partly because the ejecta are accreted onto pre-existing stars, although it requires that a large fraction of the total mass in stars with $M > 10 \, M_\odot$ becomes available to be swept up by the proto-stellar discs, and is then accreted onto the protostars. Another suggestion is that the proto-cluster gas was reprocessed and polluted by super-massive stars formed in run-away collisions during the early stages of the cluster formation \citep{Denissenkov2013}.

Clearly, a crucial piece of information is whether the GCs in dwarf galaxies resemble those in the Milky Way by actually containing multiple stellar populations, and if so, in what proportions. Of the systems mentioned above, the Fornax dSph is by far the closest, and hence the best suited for addressing this question. From observations of 9 red giants in three of the Fornax GCs (Fornax 1, 2, and 3) \citet{Letarte2006} found a hint of the Na-O anti-correlation. 
\citet{DAntona2013} used the Hubble Space Telescope (HST) photometry of \citet{Buonanno1998} to study the horizontal branch morphologies of the clusters, and suggested that the Fornax GCs contain second-generation stars in about the same proportion seen in Milky Way GCs, roughly 50\%. In this paper we use new photometry in the ultraviolet, obtained with the Wide Field Camera 3 on board the HST, to directly search for stars with anomalous chemical composition.

\section{Observations and data reduction}

\begin{figure}
\includegraphics[width=\columnwidth]{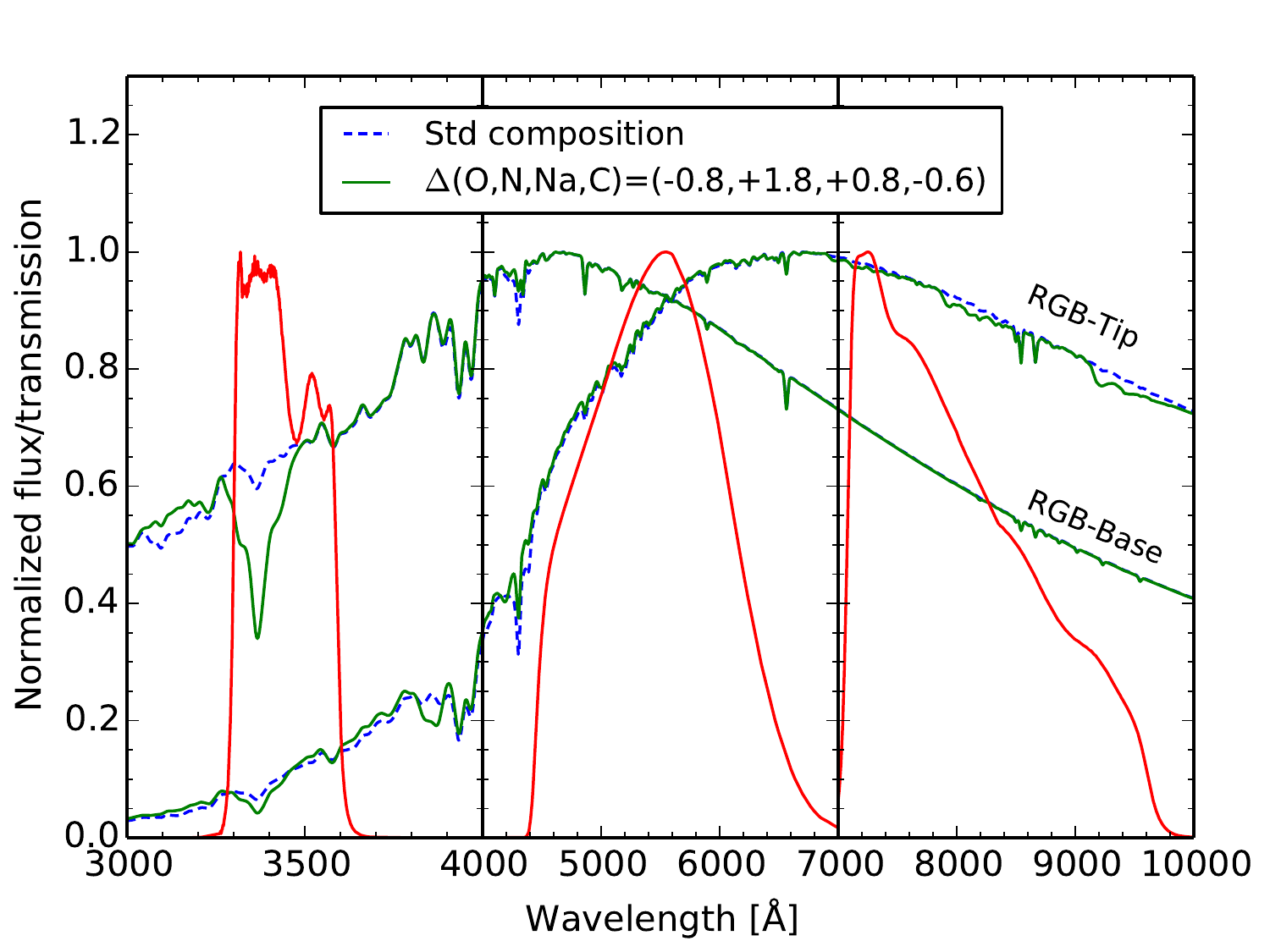}
\caption{\label{fig:filt}Model spectra and filter transmission curves (WFC3/F343N, WFPC2/F555W and WFPC2/F814W). Models are shown for $\mathrm{[Fe/H]}=-2.3$ and $[\alpha/\mathrm{Fe}]=+0.2$ for stars near the base of the RGB ($\log g = 3.2$ and $T_\mathrm{eff} = 5448$ K) and near the tip of the RGB ($\log g = 0.7$ and $T_\mathrm{eff} = 4413$ K). The  model spectra have been smoothed with a Gaussian kernel with $\sigma=10$~\AA.\\}
\end{figure}

Photometry in the blue and ultraviolet can  reveal abundance variations that affect the strengths of molecular absorption features from NH, CH and CN \citep{Hesser1977,Grundahl2002,Yong2008,Sbordone2011,Monelli2013,Kravtsov2014}. The WFC3 F343N filter is well suited for such observations as it provides maximum sensitivity to the NH band near 3370 \AA. This is illustrated in Fig.~\ref{fig:filt}, which shows the transmission curves of the F343N filter and the WFPC2 F555W and F814W filters, calculated with the \texttt{synphot} task in the \texttt{STSDAS} package in \texttt{IRAF}.
We also show model spectra for metal-poor stars at the base and tip of the red giant branch (RGB), calculated with the \texttt{ATLAS12} and \texttt{SYNTHE} codes \citep{Sbordone2004,Kurucz2005}. For both stars, spectra are shown for standard composition (with an $\alpha$-enhancement of $+0.2$ dex and metallicity $\mathrm{[Fe/H]}=-2.3$) and for the ``CNONaI'' mixture of \citet{Sbordone2011} that has a N-enhancement of $+1.8$ dex, which is typical of the most N-rich ``second-generation'' stars in Galactic GCs. For the star near the base of the RGB, the increased strength of the NH absorption band in the N-enhanced model spectrum leads to a decrease in the flux through the F343N filter of about 0.16 mag, while the F555W and F814W fluxes are practically unaffected. For stars at a fixed luminosity on the lower RGB, the F343N-F555W color is thus expected to be a sensitive indicator of the N abundance, while the F555W-F814W color is expected to be insensitive to light element abundance variations. For brighter RGB stars (at $M_V \la -1$), the sensitivity of the F343N-F555W color to N abundance gradually decreases, and for the tip-RGB model the color difference between normal and N-enhanced composition is only 0.08 mag.

We observed the four metal-poor GCs (Fornax 1, 2, 3, and 5) with the F343N filter (program ID 13295, P.I. S. Larsen). Each cluster was observed for four orbits, making use of the standard box dither pattern with a point spacing of $0\farcs173$. Within each orbit, two exposures with a dither offset of $0\farcs112$ were made.  The total integration time was 11500 s or 3 hr 11 min per cluster.
The natural sky background is very dark in F343N, which can lead to significant losses of the signal in individual pixels due to inefficiency of charge transfer during read-out, and we therefore made use of the post-flash option to increase the background level to 10 counts per pixel. To further reduce charge transfer losses, the clusters were placed as close to the read-out register as possible, in one of the quadrants of the WFC3 CCD mosaic. Finally, we used the Fortran program \texttt{wfc3uv\_ctereverse}\footnote{\texttt{http://www.stsci.edu/hst/wfc3/tools/cte\_tools}} to correct for remaining charge transfer losses. 

For Fornax 3, the guide star acquisition failed for one of the orbits. As a result, the pointing drifted significantly during one of the two exposures obtained in this orbit and a repeat observation was made. The repeat observation could not be made at the same roll angle, which slightly complicated the subsequent data reduction for Fornax 3 (see below). However, since we could still use the other exposure made during the problematic orbit, this also meant that the total useable integration time was somewhat longer (12827 sec) for this cluster.

The WFC3 observations were combined with archival WFPC2 F555W and F814W images \citep[program ID 5917, P.I. R. Zinn;][]{Buonanno1998}. These data have integration times of 5640~s in F555W and 7720~s in F814W with all four clusters roughly centered on the PC chip, which has a spatial sampling only slightly worse than that of WFC3 ($0\farcs0455$ vs. $0\farcs040$ per pixel). For the WFPC2 data, corrections for CTE losses were applied to the photometry following \citet{Dolphin2009}. In this paper we will occasionally use the letters U, V and I to refer to the F343N, F555W and F814W filters, although we will always work in the instrumental systems and it should be remembered that the F343N filter in particular is very different from a standard U filter.

We did not include the more metal-rich cluster Fornax~4 \citep[$\mathrm{[Fe/H]}\sim-1.4$;][]{Larsen2012a} in our program as there are many field stars in the Fornax dSph with metallicities similar to that of Fornax~4, making this cluster less suitable for constraining GC mass loss scenarios. Furthermore, the existing archival F555W and F814W HST imaging of Fornax~4 \citep[used by][]{Buonanno1999} is significantly shallower than the corresponding datasets for the other clusters (2400 s in each band), and has the cluster imaged on the WF3 detector with its poorer spatial sampling ($0\farcs1$ per pixel).

\begin{figure}
\begin{minipage}{42mm}
Fornax 1  \smallskip \\
\includegraphics[width=42mm]{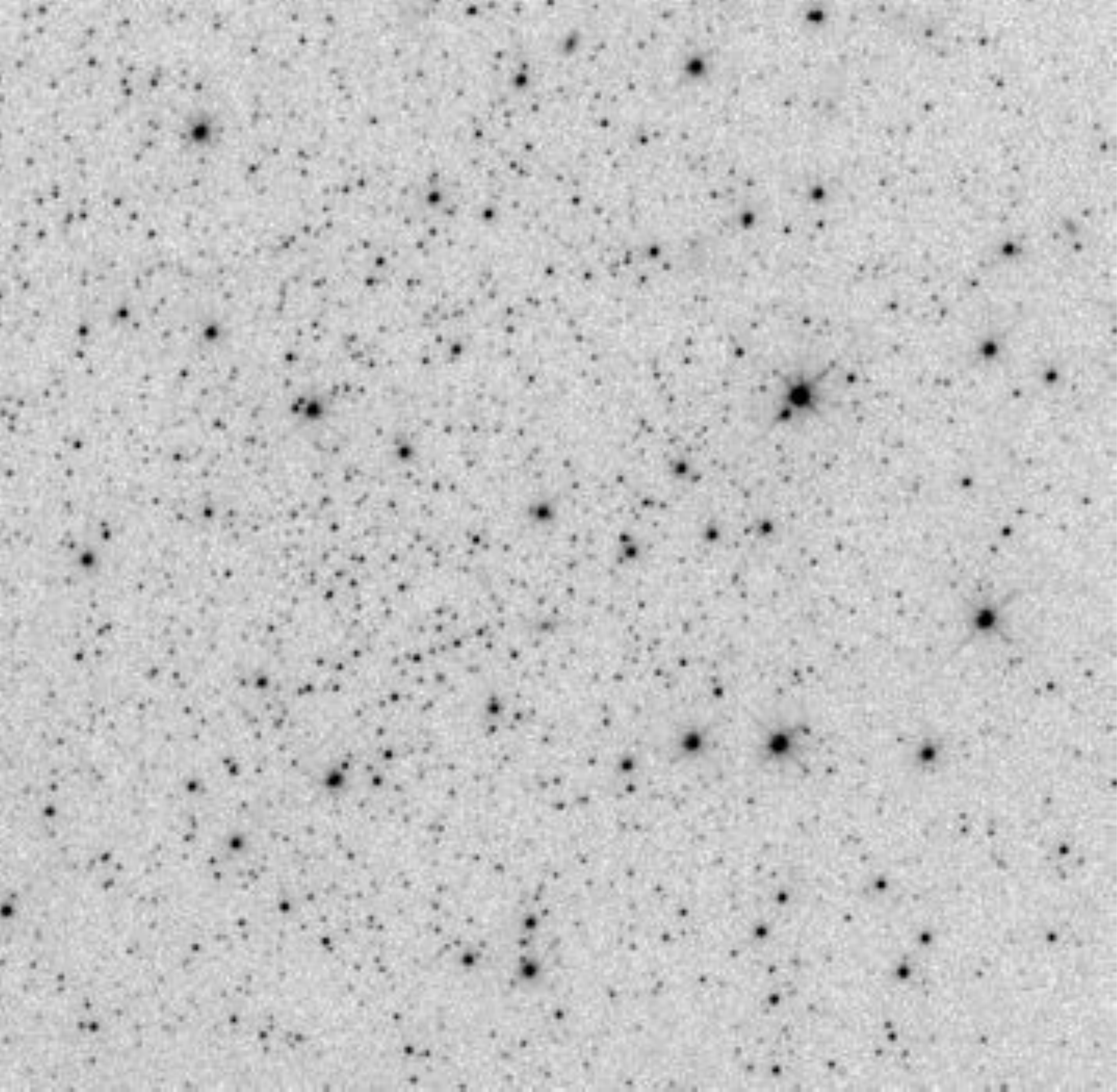}
\end{minipage} \smallskip
\begin{minipage}{42mm}
Fornax 2 \smallskip \\
\includegraphics[width=42mm]{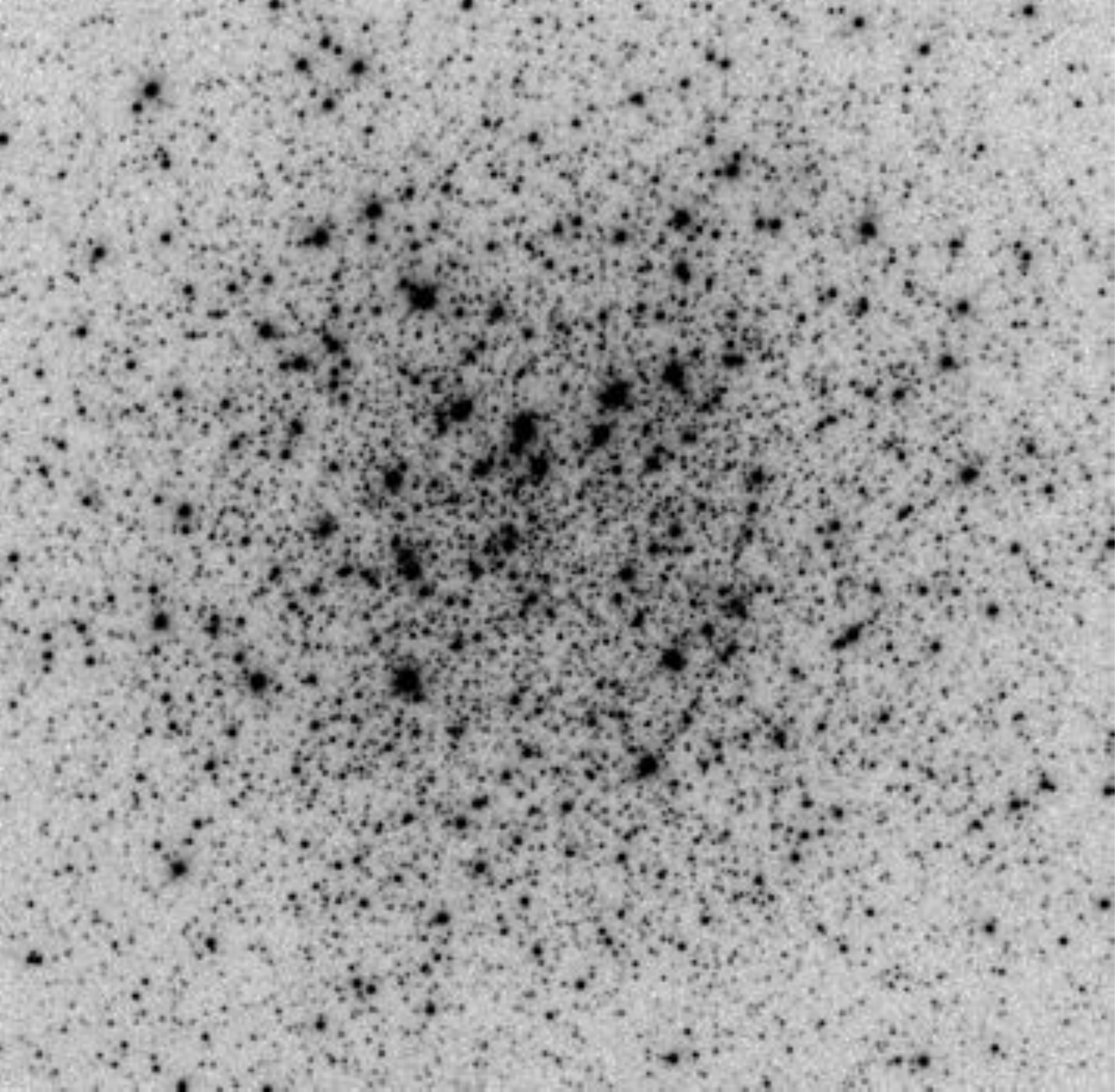}
\end{minipage} \smallskip \\
\begin{minipage}{42mm}
Fornax 3 \smallskip \\
\includegraphics[width=42mm]{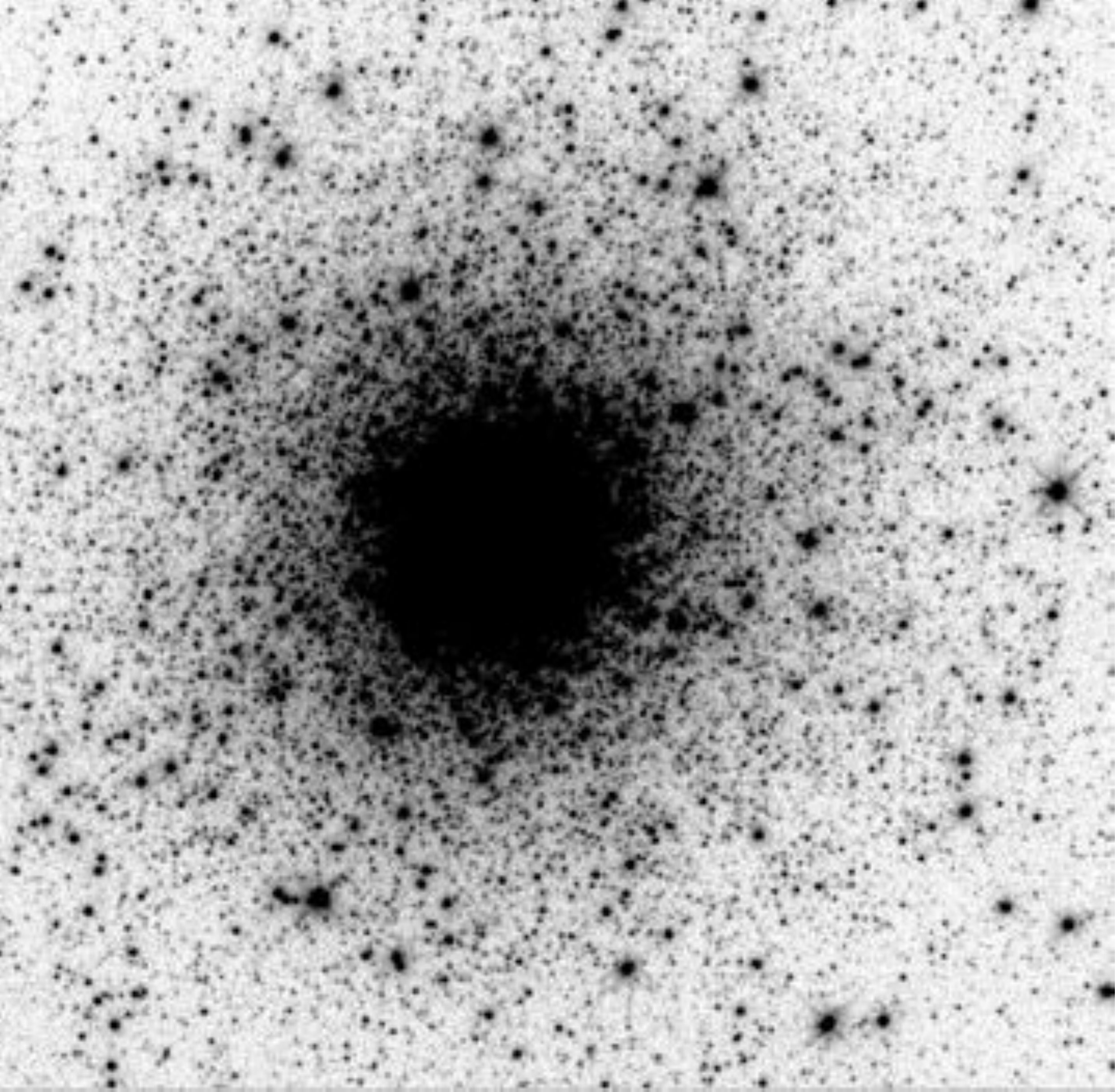}
\end{minipage}
\begin{minipage}{42mm} 
Fornax 5 \smallskip \\
\includegraphics[width=42mm]{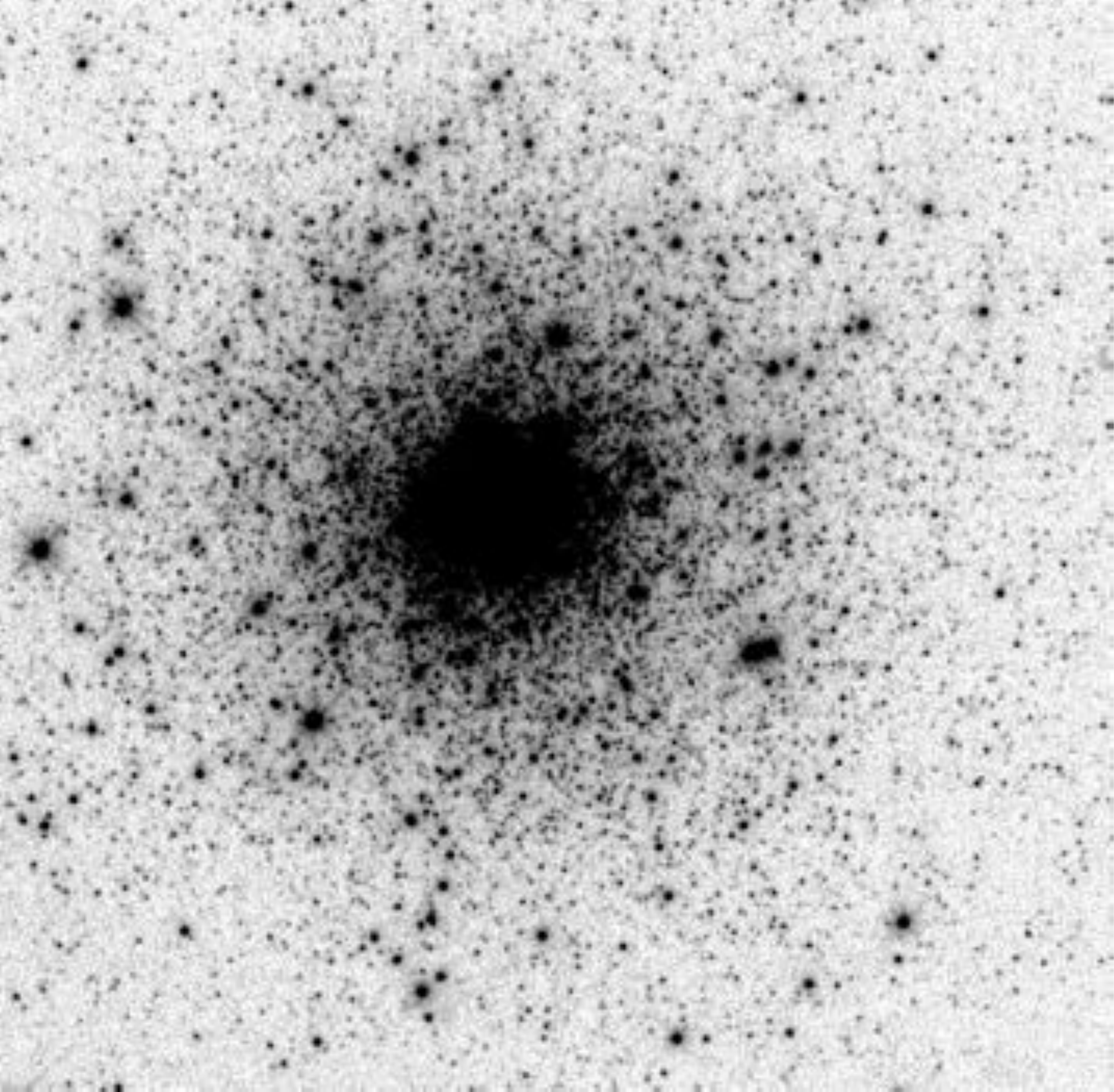}
\end{minipage}
\caption{\label{fig:images}WFPC2/PC F555W images of the clusters. For all four clusters we show the full PC field of view ($36\arcsec\times36\arcsec$ or $24\times24$ pc) with the same contrast settings, emphasizing the enormous range in central concentration.\\}
\end{figure}

The F555W images of the four metal-poor clusters are shown in Fig.~\ref{fig:images}. The same contrast settings have been used in all four panels to illustrate the enormous range in central concentration spanned by the clusters. Fornax~1 and Fornax~2 are clearly far more diffuse than Fornax~3 and Fornax~5 \citep{Mackey2003a}. We note that this is not easily explained as a simple radial trend - while Fornax~1, the most diffuse of the clusters, is indeed the one located at greatest (projected) distance from the galaxy center \citep[$0.67^\circ$;][]{Mackey2003a}, Fornax~5 is nearly as distant ($0.60^\circ$), but much more compact. In spite of the very low density of Fornax~1, the surrounding stellar density of the Fornax dwarf itself is so low that field stars still contribute negligibly to the color-magnitude diagram within the PC field of view. 

After experimenting with various ways of obtaining photometry from the datasets, we settled on the following procedure: The individual pipeline-reduced (and, in the case of F343N, CTE-corrected) images were first multiplied by the pixel-area maps provided by STScI. For each band, we then constructed a ``master frame'' by aligning the individual exposures using integer pixel shifts (thereby avoiding any rebinning) and average-combining the shifted exposures with the \texttt{imcombine} task in \texttt{IRAF}. We used the \texttt{ccdclip} option in \texttt{imcombine}  to reject cosmic rays and other artefacts (such as hot/dead pixels). The bad pixel lists produced by \texttt{imcombine} were then used to replace bad pixels in the individual images with pixels from the master frame. Point-spread function (PSF) fitting photometry was done on the cleaned individual frames with \texttt{ALLFRAME} \citep{Stetson1994}, following the usual procedure of an initial \texttt{ALLFRAME} run, re-determination of the PSF and detection of additional stars in a star-subtracted image, followed by a second iteration.
The magnitude of each star was then obtained as a weighted average of the measurements on the individual frames, taking the differences in exposure times and corresponding variations in signal-to-noise ratio into account. This procedure was found to produce far better results than carrying out the photometry on the combined images.
Furthermore, it has the added advantage of allowing an assessment of the photometric errors from the dispersion of the magnitudes measured on the individual frames. 
The PC F555W and F814W frames were all measured simultaneously in \texttt{ALLFRAME}, but the WFC3 F343N data had to be measured separately because of the different geometric distortions  in the WFPC2 and WFC3 images. In the case of Fornax~3, the two repeat observations were also measured separately and subsequently combined with the measurements on the other seven F343N exposures. The PC and WFC3 photometry was finally combined by setting up coordinate transformations between the different frames with the \texttt{geomap} task in \texttt{IRAF} and matching the star lists based on the transformed coordinates. 

For a direct comparison with a well-studied GC with a known spread in N abundance and a metallicity similar to that of the Fornax GCs, we also observed the Galactic globular cluster M15. In this cluster, a large spread (about 2 dex) in the N abundance is well established from spectroscopy of individual stars \citep{Sneden1997,Cohen2005a}. Because of the much smaller distance, only short exposures of about 90 s were needed to reach comparable depth in F343N. However, to reduce the overhead due to WFC3 buffer dumps we exposed for $2\times340$ s in F343N, leaving enough time for additional short exposures of $2\times10$ s in F555W and F814W within a single orbit. Since all observations of M15 were obtained with WFC3, we could measure the F343N, F555W and F814W magnitudes in a single \texttt{ALLFRAME} run.

The \texttt{ALLFRAME} photometry was calibrated to the STMAG system by tying it to aperture photometry of the PSF reference stars, using apertures with radii of $0\farcs4$ (for WFC3) and $0\farcs5$ (for WFPC2). For the WFC3, we used the zero-points from the WFC3 web page\footnote{\texttt{http://www.stsci.edu/hst/wfc3/phot\_zp\_lbn}}, which refer to the same aperture size used for the photometry ($Z_\mathrm{F343N} = 22.7506$ mag, $Z_\mathrm{F555W} = 25.6216$ mag, and $Z_\mathrm{F814W} = 25.8226$ mag).
For WFPC2 we used the zero-points from the WFPC2 data handbook \citep{Gonzaga2010}, which are $Z_\mathrm{F555W} = 22.545$ mag and $Z_\mathrm{F814W}=22.902$ mag. Because the WFPC2 zero-points refer to an infinite aperture, we applied an aperture correction of $-0.1$ mag to account for the $0\farcs5$ aperture used for the aperture photometry. A further offset of $-0.745$ mag was added to account for the ratio of 1.987 between the gain factors of the standard gain 7 setting and the gain 14 setting used for these WFPC2 observations. The photometry is listed in Table~\ref{tab:f1}--\ref{tab:f5}.

\begin{deluxetable*}{lccccccccccc}
\tablecaption{\label{tab:f1}Photometry for Fornax 1.}
\tablecolumns{12}
\tablehead{
 ID  & X & Y & \multicolumn{3}{c}{F555W} & \multicolumn{3}{c}{F814W} & \multicolumn{3}{c}{F343N} \\
 & & & \colhead{m}   & \colhead{err} & \colhead{rms} &
         \colhead{m}   & \colhead{err} & \colhead{rms} &
         \colhead{m}   & \colhead{err} & \colhead{rms}
}         
\tabletypesize{\small}
\startdata
   79 & 414.06 &  87.55 & 25.181 & 0.052 & 0.109 & 25.988 & 0.067 & 0.340 & 25.603 & 0.109 & 0.364 \\
   92 & 560.80 & 90.51 & 22.857 & 0.016 & 0.036 & 23.316 & 0.013 & 0.039 & 23.216 & 0.027 & 0.053 \\
  108 & 544.07 & 91.88 & 24.774 & 0.036 & 0.111 & 25.624 & 0.052 & 0.271 & 24.813 & 0.062 & 0.226 \\
  119 & 438.90 & 93.66 & 25.296 & 0.052 & 0.177 & 25.834 & 0.054 & 0.173 & 25.167 & 0.083 & 0.252 \\
  121 & 596.29  & 93.95 & 24.024 & 0.021 & 0.080 & 24.720 & 0.027 & 0.091 & 24.156 & 0.045 & 0.079 
\enddata
\tablecomments{
The columns labeled X and Y give the coordinates of the star in the F555W image. For each band, we list the measured magnitude (m), the corresponding error from the \texttt{ALLFRAME} photometry (err) and the rms of the individual measurements (rms).
Table 1 is published in its entirety in the electronic edition of the Astrophysical Journal, a portion is shown here for guidance regarding its form and content.
}
\end{deluxetable*}

\begin{deluxetable*}{lccccccccccc}
\tablecaption{\label{tab:f2}Photometry for Fornax 2.}
\tablecolumns{12}
\tablehead{
 ID  & X & Y & \multicolumn{3}{c}{F555W} & \multicolumn{3}{c}{F814W} & \multicolumn{3}{c}{F343N} \\
 & & & \colhead{m}   & \colhead{err} & \colhead{rms} &
         \colhead{m}   & \colhead{err} & \colhead{rms} &
         \colhead{m}   & \colhead{err} & \colhead{rms}
}         
\tabletypesize{\small}
\startdata
  801 & 334.94 & 154.97 & 23.192 & 0.013 & 0.050 & 23.644 & 0.015 & 0.060 & 23.760 & 0.034 & 0.138 \\
  809 & 176.28 & 155.52 & 25.024 & 0.044 & 0.147 & 25.758 & 0.055 & 0.132 & 25.115 & 0.083 & 0.442 \\
  810 & 288.50 & 155.97 & 23.561 & 0.016 & 0.039 & 24.002 & 0.015 & 0.038 & 23.890 & 0.038 & 0.061 \\
  811 & 375.46 & 156.55 & 26.197 & 0.121 & 0.319 & 26.631 & 0.106 & 0.435 & 27.093 & 0.361 & 1.259 \\
  812 & 384.12 & 156.47 & 24.549 & 0.030 & 0.097 & 25.253 & 0.032 & 0.091 & 24.539 & 0.056 & 0.128 
\enddata
\tablecomments{
Table 2 is published in its entirety in the electronic edition of the Astrophysical Journal, a portion is shown here for guidance regarding its form and content.
}
\end{deluxetable*}

\begin{deluxetable*}{lccccccccccc}
\tablecaption{\label{tab:f3}Photometry for Fornax 3.}
\tablecolumns{12}
\tablehead{
 ID  & X & Y & \multicolumn{3}{c}{F555W} & \multicolumn{3}{c}{F814W} & \multicolumn{3}{c}{F343N} \\
 & & & \colhead{m}   & \colhead{err} & \colhead{rms} &
         \colhead{m}   & \colhead{err} & \colhead{rms} &
         \colhead{m}   & \colhead{err} & \colhead{rms}
}         
\tabletypesize{\small}
\startdata
   21 & 134.54 & 78.93 & 22.973 & 0.022 & 0.100 & 23.395 & 0.019 & 0.141 & 23.352 & 0.030 & 0.070 \\
   23 & 358.31 & 78.29 & 25.201 & 0.057 & 0.159 & 25.755 & 0.063 & 0.209 & 25.377 & 0.085 & 0.245 \\
   24 & 398.39 & 78.83 & 24.939 & 0.049 & 0.095 & 25.746 & 0.062 & 0.230 & 25.122 & 0.082 & 0.252 \\
   25 & 451.88 & 78.52 & 23.046 & 0.018 & 0.126 & 23.470 & 0.017 & 0.097 & 23.410 & 0.028 & 0.087 \\
   32 & 402.25 & 79.66 & 25.549 & 0.066 & 0.164 & 26.194 & 0.090 & 0.604 & 25.333 & 0.092 & 0.092 
\enddata
\tablecomments{
Table 3 is published in its entirety in the electronic edition of the Astrophysical Journal, a portion is shown here for guidance regarding its form and content.
}
\end{deluxetable*}

\begin{deluxetable*}{lccccccccccc}
\tablecaption{\label{tab:f5}Photometry for Fornax 5.}
\tablecolumns{12}
\tablehead{
 ID  & X & Y & \multicolumn{3}{c}{F555W} & \multicolumn{3}{c}{F814W} & \multicolumn{3}{c}{F343N} \\
 & & & \colhead{m}   & \colhead{err} & \colhead{rms} &
         \colhead{m}   & \colhead{err} & \colhead{rms} &
         \colhead{m}   & \colhead{err} & \colhead{rms}
}         
\tabletypesize{\small}
\startdata
  225 & 501.73 & 123.66 & 24.750 & 0.038 & 0.110 & 25.504 & 0.048 & 0.163 & 24.728 & 0.066 & 0.174 \\
  227 & 144.89 & 125.51 & 25.427 & 0.061 & 0.187 & 25.969 & 0.062 & 0.207 & 25.473 & 0.095 & 0.316 \\
  230 & 338.39 & 125.03 & 22.200 & 0.009 & 0.020 & 22.563 & 0.011 & 0.024 & 22.747 & 0.024 & 0.078 \\
  233 & 466.55 & 125.45 & 25.752 & 0.072 & 0.326 & 26.448 & 0.099 & 0.649 & 25.804 & 0.129 & 0.300 \\
  234 & 614.64 & 124.63 & 25.207 & 0.052 & 0.223 & 25.867 & 0.058 & 0.161 & 25.234 & 0.082 & 0.266 
\enddata
\tablecomments{
Table 4 is published in its entirety in the electronic edition of the Astrophysical Journal, a portion is shown here for guidance regarding its form and content.
}
\end{deluxetable*}

The photometry was corrected for foreground reddening using the \citet{Schlafly2011} values (via NED, the \textit{NASA/IPAC Extragalactic Database}). These are $E(V\!-\!I)=0.022$ mag, 0.039 mag, 0.031 mag, and 0.027 mag for Fornax 1, 2, 3, and 5, respectively, and $E(V\!-\!I)=0.135$ mag for M15 (here, $V$ and $I$ are the Landolt filters). To find the extinction in the HST bands we used the extinction curve of \citet{Cardelli1989}.
Throughout this paper we assume $(m-M)_0 = 20.68$ mag for all four Fornax GCs \citep{Buonanno1998}, although the depth of the Fornax GC system may imply a range of $\sim0.15$ mag for the distance moduli \citep{Mackey2003}. For M15 we assume $(m-M)_0 = 15.06$ mag \citep{VandenBosch2006}.

\section{Results}

\begin{figure*}
\includegraphics[width=\textwidth]{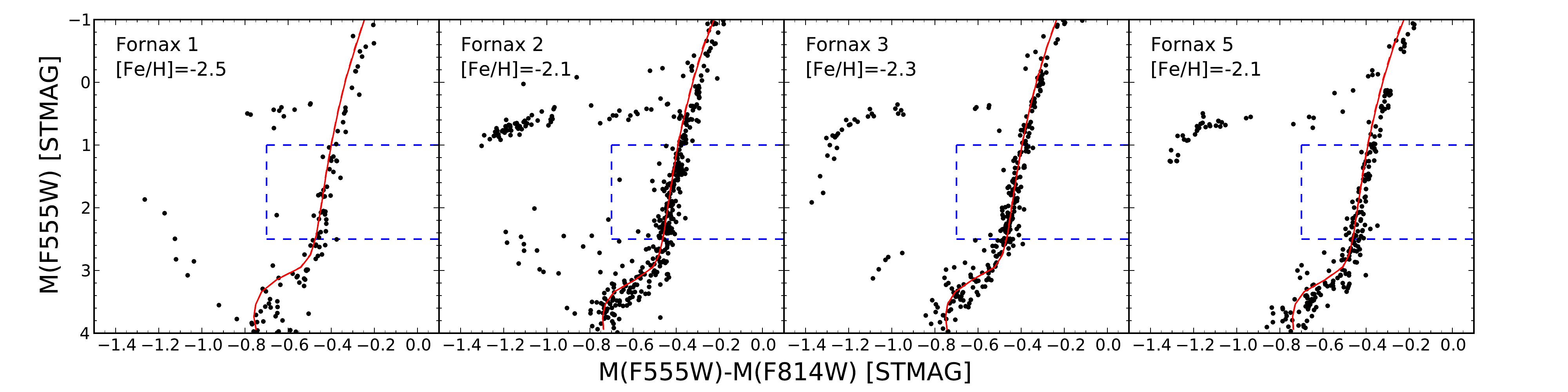}
\includegraphics[width=\textwidth]{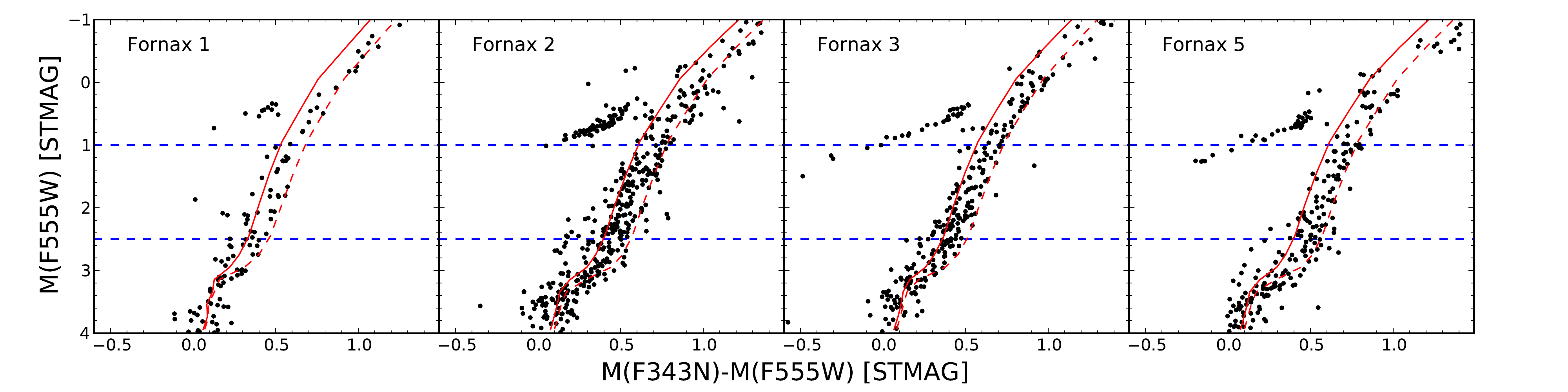}
\caption{\label{fig:vif}$M_\mathrm{F555W}$ vs.\ $(M_\mathrm{F555W}-M_\mathrm{F814W})$ (top) and
$M_\mathrm{F555W}$ vs.\
$(M_\mathrm{F343N}-M_\mathrm{F555W})$  (bottom) CMDs for the Fornax GCs. 
Isochrones for ages of 13 Gyr and the metallicities indicated in the upper panels are shown as red curves. The dashed red curves are for the N-enhanced ``CNONaI'' mixture \citep{Sbordone2011}.
Stars are selected within the radial ranges indicated in Table~\ref{tab:spread} and the blue dashed lines indicate the magnitude and color ranges used to measure the color spreads.
}
\end{figure*}

\begin{figure}
\includegraphics[width=\columnwidth]{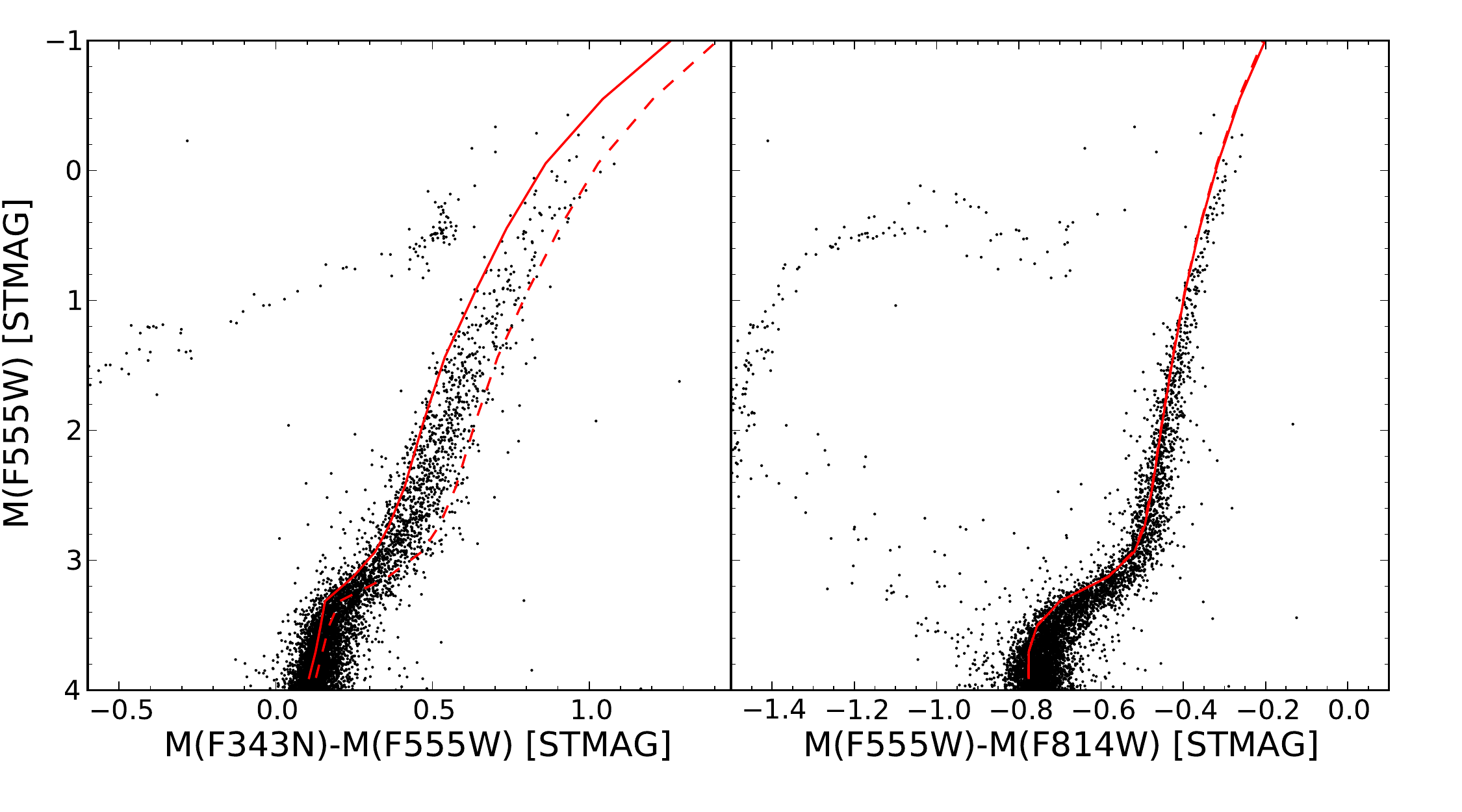}
\caption{\label{fig:m15}Color-magnitude diagrams for M15. The red curves show \texttt{ATLAS12}/\texttt{SYNTHE} model colors for a Dotter isochrone with $\mathrm{[Fe/H]}=-2.2$ and $[\alpha/\mathrm{Fe}]=+0.4$. The solid and dashed line styles indicate standard and N-enhanced composition, respectively, as in Fig.~\ref{fig:vif}.}
\end{figure}

The color-magnitude diagrams (CMDs) for the Fornax GCs and M15 are shown in Fig.~\ref{fig:vif} and Fig.~\ref{fig:m15}, respectively. We have excluded stars in the crowded central regions of the clusters, except for Fornax 1 which is so diffuse that stars can easily be resolved and measured all the way to the center. The adopted inner radii are given in Table~\ref{tab:spread} and represent a compromise between maximizing the number of stars and keeping the photometric errors small. For Fornax~3 and Fornax~5, our inner radii are the same as those adopted by \citet{Buonanno1998} for their ``faint sample'' (i.e., $V>22$), while we found that we could obtain good photometry for stars somewhat closer to the center in Fornax~2.
Further, we have only included stars for which the rms deviation of the individual F343N magnitude measurements is rms$_\mathrm{F343N} < 0.1$ mag, corresponding to a formal random error of $\sigma_\mathrm{F343N} < 0.04$ mag on the average combined F343N magnitudes. 
The (F555W-F814W, F555W) CMDs in the top row of Fig.~\ref{fig:vif} are very similar to the corresponding CMDs, based on the same data, published by \citet{Buonanno1998}. We note that our selection on rms$_\mathrm{F343N}$ causes some incompleteness on the horizontal branch in the color range $-0.9\la \mathrm{F555W-F814W} \la -0.6$, where RR Lyrae stars exhibit significant variability on time scales similar to those over which our observations were carried out \citep{Mackey2003,Greco2009}.
For reference, the metallicities of the Fornax GCs derived from high-dispersion spectroscopy \citep{Letarte2006,Larsen2012a} are indicated in the upper panels of Fig.~\ref{fig:vif}. 

We also include model colors for standard RGB stars and stars with the CNONaI mixture,  computed for the isochrones of \citet{Dotter2007} for the metallicity of each cluster. We assume an age of 13 Gyr in all cases. The model colors for the metallicities and [$\alpha$/Fe] ratios used in this paper are listed in Table~\ref{tab:modcol}.
While the light element abundance variations are not taken into account in the isochrones, these variations have little effect on the isochrones themselves \citep{Sbordone2011}.  As expected, the model F555W-F814W colors are virtually independent of the light element abundances, while the F343N-F555W colors of the N-normal and N-enhanced models differ by about 0.16 mag. The model colors agree quite well with the observed CMDs, supporting the low metallicities derived from high-dispersion spectroscopy, and the red giants in both the Fornax GCs and M15 tend to fall between the N-normal and N-enhanced models. 
The exact relative locations of the data and models in these plots are, of course, sensitive to our assumptions about age, metallicity, $[\alpha/\mathrm{Fe}]$ ratios, reddening, and distances of individual clusters.
In addition, the calculation of model colors from the physical properties ($T_\mathrm{eff}$, $\log g$, chemical composition) is also uncertain and dependent on the specific model atmospheres used \citep{Cassisi2004,Worthey2011}, the completeness of the line lists used in the calculation of synthetic spectra, etc. 
We note that the \texttt{ATLAS12} atmospheres used here were computed for the specific abundance patterns used for the synthetic spectra. 
However, our main aim here is not to carry out a detailed comparison of the models and data, but primarily to quantify the spread in the colors of the RGB stars.

\begin{figure}
\includegraphics[width=90mm]{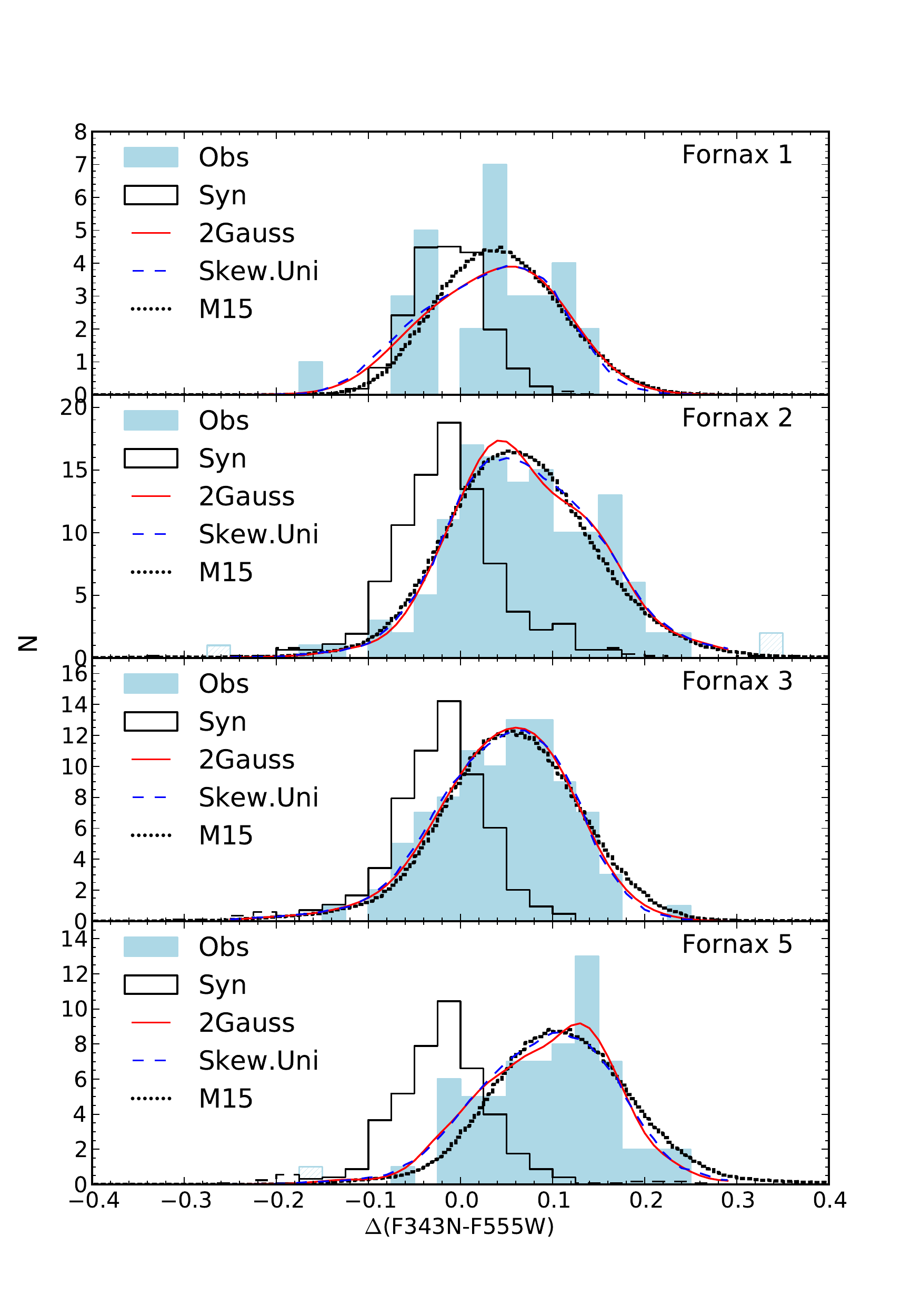}
\caption{\label{fig:uvhist}Observed and simulated $\Delta$(F343N-F555W) color distributions. The hashed/dashed parts of the histograms indicate data values outside the 3$\sigma$ limits that were excluded when computing the color spreads in Table~\ref{tab:spread}.
The red curves are double-Gaussian models convolved with the error distributions, while the blue dashed curves are model curves for skewed uniform distributions (Section~\ref{sec:mcmc}-\ref{sec:umcmc} and Table~\ref{tab:gfits}-\ref{tab:ufits}). The dotted curves show the M15 $\Delta$(F343N-F555W) color distribution convolved with the error distributions.\\
}
\end{figure}

\subsection{Quantification of color spreads: artificial star tests}

\label{sec:cspreads}

The first impression from the CMDs is that the observed F343N-F555W colors show a considerably larger spread on the RGB than the F555W-F814W colors in both the Fornax GCs and in M15. 
To quantify the contribution of photometric errors to the observed spreads, we carried out artificial star experiments. For each Fornax cluster, we started by selecting RGB stars with $m_\mathrm{F555W}<24$ from the photometry files. For each star, we then interpolated in an isochrone to obtain synthetic F343N-F555W and F555W-F814W colors at the corresponding F555W magnitude. This gave us a list of stars with a magnitude distribution similar to that of real RGB stars, but with no intrinsic spread in color at a given magnitude. We then generated coordinate lists for the artificial stars, each containing 100 stars in each of the radial bins $4\arcsec-6\arcsec$, $6\arcsec-8\arcsec$, $8\arcsec-10\arcsec$ and $10\arcsec-12\arcsec$. 
The artificial stars were required to have a minimum separation of at least 20 pixels in order to avoid introducing artificial crowding. Four to six such coordinate lists were generated for each radial bin, for a total of about 2000 artificial stars per cluster. An entry from the list of magnitudes and colors was assigned to each coordinate and the artificial stars were then added, 100 at a time, to the images, using the \texttt{mksynth} task in the \texttt{BAOLAB} package \citep{Larsen1999}. We also added 15--20 extra artificial stars at relatively isolated locations with magnitudes similar to those of the PSF stars used for the \texttt{ALLFRAME} photometry ($m_\mathrm{F343N}\approx21.5$). These stars were used as PSF stars for the artificial star tests, so that these tests also took into account the uncertainties involved in (re-)determining the PSFs. 
The photometry procedures were then repeated, including the selection on rms and radial coordinate.

In Table~\ref{tab:spread} and in Figure~\ref{fig:uvhist} we compare the color spreads for the cluster stars with the artificial star experiments for RGB stars with $+1 < M_\mathrm{F555W} < +2.5$ (the magnitude range is indicated by the horizontal dashed lines in Fig.~\ref{fig:vif}).
This range avoids confusion with horizontal branch stars and excludes brighter RGB stars where deep mixing may have brought processed material to the surface.  In metal-poor field giants, the signatures of internal mixing appear for luminosities $\log L/L_\odot \ga 1.8$
\citep{Gratton2000} or $M_\mathrm{F555W}\approx M_V\la+0.5$, well above our adopted limit.
We have further imposed a color cut of F555W-F814W $>-0.7$ to exclude blue stragglers and extreme horizontal branch stars, and we have applied the same selection on rms$_\mathrm{F343N}$ and distance from the cluster centers as in Figure~\ref{fig:vif}.
We denote by $\Delta$(F555W-F814W) the difference between the measured F555W-F814W color and an isochrone of the appropriate metallicity (using N-normal model colors), and similarly for F343N-F555W. When computing the dispersions in Table~\ref{tab:spread}, we excluded outliers deviating by more than 3 $\sigma$ from the mean values (iterating three times). The (few) stars affected by this cut are indicated by hashed/dashed histogram styles in Figure~\ref{fig:uvhist}.

From Table~\ref{tab:spread}, the observed $\Delta$(F555W-F814W) color dispersions ($\sigma_\mathrm{obs,VI}$) are generally quite small ($\approx0.025$ mag) and very well reproduced by the artificial star experiments ($\sigma_\mathrm{synt,VI}$). This is consistent with the expectation that the F555W-F814W colors should not exhibit any intrinsic spread (in the absence of overall metallicity variations).
The observed $\Delta$(F343N-F555W) dispersions ($\sigma_\mathrm{obs,UV}$) are, in contrast, significantly larger than the dispersions in the artificial star experiments ($\sigma_\mathrm{synt,UV}$). 
The $P$ values from a Levene test for similar variances confirm that the differences between $\sigma_\mathrm{obs,UV}$ and $\sigma_\mathrm{synt,UV}$ are, in all cases, highly significant ($P<10^{-5}$), while the $P$ values for $\Delta$(F555W-F814W) do not indicate any significant differences between the observed and simulated dispersions. It is also evident from Fig.~\ref{fig:uvhist} that the observed $\Delta$(F343N-F555W) distributions are significantly broader than those seen in the artificial star tests.

Subtracting the $\Delta$(F343N-F555W) dispersions of the artificial star tests from those of the observed distributions in quadrature, $\sigma_{\mathrm{true,UV}}^2 = \sigma_\mathrm{obs,UV}^2 - \sigma_\mathrm{synt,UV}^2$, we find
$\sigma_{\mathrm{true,UV}}$ = 0.061 mag (Fornax~1), 0.052 mag (Fornax~2), 0.048 mag (Fornax~3), and 0.049 mag (Fornax~5). These values are not very sensitive to the exact selection criteria. More restrictive cuts on the errors, for example, tend to make the observed dispersions smaller so that the intrinsic spreads account for a larger fraction of the total spread, but this also decreases the number of stars. 
However, because the error distributions are comparable in width to the intrinsic dispersions, the precise shapes of the intrinsic $\Delta$(F343N-F555W) distributions are poorly constrained. For a Gaussian distribution, a dispersion of $\sigma = 0.050$ mag corresponds to a FWHM of 0.12 mag. For a uniform distribution of width $w_u$, one finds $w_u = \sigma \sqrt{12}$, so a mean dispersion of $\sigma=0.050$ mag then corresponds to  $w_u \approx 0.17$ mag. These values are comparable to the 0.16 mag separation of the standard vs.\ N-enhanced model isochrones for RGB stars. There are, of course, many other possibilities. For example, if the intrinsic distributions consist of two $\delta$-functions then a  dispersion of 0.050 mag would correspond to a separation of 0.1 mag between the two peaks. We explore a few possibilities in more detail below (Sect.~\ref{sec:mcmc}-\ref{sec:umcmc}).

The constant number of artificial stars per radial bin corresponds to a surface density that depends on the radius as $1/R$. This is comparable to the slopes of the \citet{Mackey2003a} model fits to the cluster profiles near the core radius, but shallower at larger radii. There is thus a slight overrepresentation of artificial stars at larger radii in Fornax~3 and Fornax~5 (where we measure stars outside $\sim3$ core radii), compared to the real clusters. We investigated the effect of this difference on the measured dispersions of the artificial stars by applying weights to the artificial stars, computed as
\begin{equation}
w_i \propto R_i \left(1 + \frac{R_i^2}{a^2} \right)^{-\gamma/2},
\end{equation}
i.e., the weight of the $i$the star is the ratio of an \citet{Elson1987}-type profile and the $1/R$ profile of the artificial star distribution at radial coordinate $R_i$. The parameters $a$ and $\gamma$ were taken from \citet{Mackey2003a}. 
We then recomputed the dispersions, now assigning a weight $w_i$ to each artificial star.
Compared to the unweighted $\sigma_\mathrm{synt,UV}$ values listed in Table~\ref{tab:spread}, the differences were very small. For $\Delta$(F343N-F555W), Fornax~3 and Fornax~5 changed from $\sigma_\mathrm{synt,UV}=0.048$ mag to $\sigma_\mathrm{synt,UV}=0.050$ mag. For all other color distributions, the changes were 0.01 mag or less.  For our further analysis, we thus proceeded using the ``raw'' artificial star results.

As a further test of the reality of the spread in $\Delta$(F343N-F555W), we compared the color spreads of the artificial star tests and the observations at fainter magnitudes, where the model colors become less sensitive to light element abundance variations. While the color difference between the N-normal and N-enhanced models is roughly constant for $-1 < M_\mathrm{F555W}<+3$, it is only 0.03 mag at $M_\mathrm{F555W} = +3.3$. The comparison is complicated by the increase in the overall photometric errors and decreasing completeness at fainter magnitudes, but for the range $+3.2 < M_\mathrm{F555W} < +3.6$ (now increasing the allowed rms deviation to rms$_\mathrm{F343N} < 0.2$ mag) we found observed dispersions of
$\sigma_\mathrm{obs,UV} = 0.070$ mag, 0.104 mag, 0.084 mag, and 0.091 mag for Fornax 1, 2, 3, and 5, respectively. The straight average is $\langle\sigma_\mathrm{obs,UV}\rangle = 0.087\pm0.007$ mag. For the artificial star tests we found $\sigma_\mathrm{synt,UV} = 0.067$ mag, 0.092 mag, 0.121 mag, and 0.080 mag, with an average of $\langle\sigma_\mathrm{synt,UV}\rangle=0.090\pm0.012$ mag, very similar to the mean observed dispersion. For Fornax~1, 2, and 5, the Levene test for similar variances returns $P>0.1$ (for Fornax~3, $P=0.01$), indicating no significant differences between the observations and artificial star tests. However, we note that the numbers of recovered artificial stars are a factor of 5--10 lower at these faint magnitudes compared to those in Table~\ref{tab:spread}, and the close agreement between the average $\sigma_\mathrm{synt,UV}$ and $\sigma_\mathrm{obs,UV}$ must be considered somewhat fortuitous.

In the case of Fornax~2, the $\Delta$(F343N-F555W) dispersion is larger than for the other clusters. This is probably related to a poorer focus of the F343N observations for this cluster: according to the STScI focus model\footnote{\texttt{http://focustool.stsci.edu}}, the WFC3 focus deviated by about 6 $\mu$m on average from the nominal value during the Fornax 2 observations, compared to 3--4 $\mu$m for our other observations.
Indeed, the PSFs of many of the individual Fornax~2 exposures are noticeably broader than for the other clusters. For example, in the individual F343N images of Fornax~1 we typically measure FWHM values of 1.6--1.8 pixels for individual stars (using the \texttt{imexamine} task in \texttt{IRAF}). For Fornax~2, the FWHM values are 2.0--2.5 pixels. However, it is worth noting that this increase in the photometric errors is well captured by the artificial star tests.

Finally, the artificial star tests allowed us to verify the photometric calibration. Because we used the appropriate photometric zero-points to calculate the count rates for the artificial stars, we would in principle expect the $\Delta$(F343N-F555W) distributions of the artificial stars to be centered around zero. Figure~\ref{fig:uvhist} shows that this is not exactly the case; there are small offsets (between $-0.013$ mag and $-0.020$ mag).
These may be taken as an indication of the systematic uncertainties on the calibration of the PSF photometry to the standard system. The offsets are most likely related to uncertainties in the determination of the sky background for the aperture photometry of the reference stars; we note that the PSF stars are relatively faint in the F343N band. In F555W-F814W, where the PSF stars are brighter, the offsets are very small ($<0.005$ mag). By making the artificial PSF stars brighter in F343N ($m_\mathrm{F343N}=20$) we could largely eliminate the offsets also for the F343N-F555W color. 
For the real data we have no such option, of course, and we retained the artificial star tests with the fainter PSF stars in order to ensure the most realistic comparison possible. While a better calibration might be possible by carrying out a more sophisticated curve-of-growth analysis, these small uncertainties are of little consequence for our purpose and we did not pursue the matter further.

\begin{deluxetable*}{lccccccccc}
\tablecaption{\label{tab:spread}Observed and simulated color spreads for RGB stars.}
\tablecolumns{10}
\tablehead{
 &   &  &  & \multicolumn{3}{c}{$\Delta$(F555W-F814W)} & \multicolumn{3}{c}{$\Delta$(F343N-F555W)} \\
\colhead{Cluster}   & \colhead{$R_\mathrm{min}$} & \colhead{$N_\mathrm{obs}$} & \colhead{$N_\mathrm{syn}$} & $\sigma_\mathrm{obs,VI}$ & $\sigma_\mathrm{synt,VI}$ & $P$
            & $\sigma_\mathrm{obs,UV}$ & $\sigma_\mathrm{synt,UV}$ & $P$ 
}
\startdata
Fornax 1 & 0 & 30 & 777 & 0.022 & 0.022 & 0.67 & 0.072 & 0.038 & $1.8\times10^{-8}$   \\
Fornax 2 & $4\farcs5$ &  131 & 544 & 0.025 & 0.023 & 0.58 & 0.078 & 0.058 & $8.0\times10^{-7}$   \\
Fornax 3 & $6\farcs0$ &  91 & 512 & 0.023 & 0.025 & 0.25 & 0.068 & 0.048 & $1.8\times10^{-6}$   \\
Fornax 5 & $6\farcs0$ & 66 & 552 & 0.023 & 0.024 & 0.88 & 0.069 & 0.048 & $3.5\times10^{-6}$
\enddata
\tablecomments{
$N_\mathrm{obs}$ and $N_\mathrm{syn}$ are the numbers of observed and artificial stars used to compute the dispersions. The $P$ values refer to the null hypothesis that the color dispersions of the observations and artificial star measurements are the same.
}
\end{deluxetable*}

\subsection{N-normal vs. N-enhanced stars: double-Gaussian fits to the color distributions}
\label{sec:mcmc}

\begin{deluxetable}{lcccc}
\tablecaption{\label{tab:gfits}Parameters for double-Gaussian fits.}
\tablehead{
          & \colhead{$c_1$} & \colhead{$c_2 - c_1$} & \colhead{$\sigma$} & \colhead{$w_1$}
}
\startdata
Fornax 1 & $-0.011^{+0.039}_{-0.050}$ & $0.098^{+0.040}_{-0.046}$ & $0.038^{+0.021}_{-0.025}$ & $0.39^{+0.31}_{-0.24}$ \\
Fornax 2 & \phs$0.051^{+0.017}_{-0.012}$ & $0.096^{+0.019}_{-0.039}$ & $0.025^{+0.024}_{-0.013}$ & $0.65^{+0.14}_{-0.16}$ \\
Fornax 3 & \phs$0.027^{+0.032}_{-0.028}$ & $0.078^{+0.023}_{-0.041}$ & $0.031^{+0.015}_{-0.016}$ & $0.43^{+0.32}_{-0.21}$ \\
Fornax 5 & \phs$0.053^{+0.035}_{-0.021}$ & $0.099^{+0.020}_{-0.035}$ & $0.018^{+0.025}_{-0.008}$ & $0.37^{+0.15}_{-0.13}$ 
\enddata
\tablecomments{
$c_1$ is the centroid in $\Delta$(F343N-F555W) of the first Gaussian component, $c_2-c_1$ the separation between the two Gaussians, $\sigma$ the dispersion (common to both components), and $w_1$ the weight of the first Gaussian.
See Section~\ref{sec:mcmc} for details.\\}
\end{deluxetable}

To quantify the relative numbers of N-normal and N-enhanced stars, we modeled the color distributions as sums of two Gaussian functions, convolved with the error distributions as determined from the artificial star tests. We  adjusted the centroids, dispersions, and weights of the two Gaussians until the best fits to the observed $\Delta$(F343N-F555W) distributions were obtained. We emphasize that this is not meant to imply that the intrinsic color distributions necessarily consist of two distinct peaks. This may, indeed, be unlikely as judged from a comparison with Milky Way GCs, which display a bewildering complexity of color distributions on the RGB \citep{Lardo2010,Milone2013,Monelli2013}. However, this parameterisation provides a convenient way to quantify whether the color distributions are strongly skewed in one direction. 

Model color distributions, $\mathcal{M}_{2G}(\Delta \mathrm{UV})$ were calculated as
\begin{equation}
\begin{split}
  \mathcal{M}_{2G}(\Delta\mathrm{UV} | c_1, c_2, \sigma_1, \sigma_2, w_1) = \\
  \eta \sum_{i=1}^{n_\mathrm{synt}} \left[ w_1 \mathcal{G}_1(\Delta\mathrm{UV} | \Delta\mathrm{UV}_{\mathrm{syn},i} + c_1, \sigma_1)\right. \\
  \left. + (1-w_1) \mathcal{G}_2(\Delta\mathrm{UV} | \Delta\mathrm{UV}_{\mathrm{syn},i} + c_2, \sigma_2) \right]
\end{split}
\end{equation}
where, for brevity, we use $\Delta\mathrm{UV}$ for $\Delta(\mathrm{F343N}-\mathrm{F555W})$ and $\Delta\mathrm{UV}_{\mathrm{syn},i}$ is the measured color offset of the $i$th synthetic star. $\mathcal{G}_1$ and $\mathcal{G}_2$ are then the Gaussian functions centered at $\Delta\mathrm{UV}_{\mathrm{syn},i}+c_1$ and $\Delta\mathrm{UV}_{\mathrm{syn},i} + c_2$ and with dispersions $\sigma_1$ and $\sigma_2$ and the weights are  $w_1$ (with $0\leq w_1 \leq1$) and $1-w_1$. The constant $\eta$ normalizes the model distribution to unity, and we have implicitly made use of the fact that convolution is commutative. We solved for the parameters of the Gaussians by maximizing the likelihood function
\begin{equation}
  \log \mathcal{L}(\Delta\mathrm{UV}_\mathrm{obs} | c_1, c_2, \sigma_1, \sigma_2, w_1) = \sum_{i=1}^{n_\mathrm{obs}} \log \mathcal{M}_{2G}(\Delta\mathrm{UV}_{\mathrm{obs},i})
\end{equation}
where $\Delta\mathrm{UV}_\mathrm{obs}$ are the observed $\Delta$(F343N-F555W) color offsets. In practice, we kept the two Gaussian dispersions equal, $\sigma_1 = \sigma_2$, and to ensure a smooth model distribution both were required to have $\sigma > 0.007$ mag. We also required $c_2 > c_1$, so that $w_1$ is always the weight of the bluest component.

We used the \texttt{emcee} Markov Chain Monte Carlo code \citep{Foreman-Mackey2013} to sample the likelihood function over the $\sigma, c_1, c_2, w_1$ parameter space. 
A summary of the results is given in Table~\ref{tab:gfits}, which lists the median values (50\% percentiles of the MCMC samples) and uncertainty intervals corresponding to the 16\% and 84\% percentiles (thus roughly equivalent to 1$\sigma$ errors) for each parameter. The model color distributions corresponding to the median parameter values are shown as smooth (red) curves in Fig.~\ref{fig:uvhist}. 
We see that the best-fitting separations of the two Gaussians are close to 0.1 mag, as already anticipated above. There is some degeneracy between the dispersions and separation of the two Gaussians; a smaller separation can be compensated by larger dispersions. A limiting case would be a single, very broad Gaussian.
However, here we are mainly concerned with the weights, $w_1$. We see that solutions with two Gaussians of roughly equal weights are preferred in all clusters. If we associate the two Gaussian peaks with ``first'' and ``second'' generation stars, then about 40\% of the stars belong to the first generation in Fornax 1, 3, and 5, while Fornax~2 has about 60\% first-generation stars. These numbers are consistent with the impression from Fig.~\ref{fig:uvhist} that the $\Delta$(F343N-F555W) distribution of Fornax~2 appears somewhat more skewed towards the left, whereas Fornax~3 and Fornax~5 are more skewed towards the right.  Fornax~1 has too few stars to provide a meaningful constraint on the exact ratio, although the spread appears to be comparable to that in the other clusters. Nevertheless, when the uncertainties are taken into account, all clusters are consistent with equal numbers of first- and second generation stars.

\subsection{Fitting skewed uniform distributions to the color distributions}
\label{sec:umcmc}

\begin{deluxetable}{lccc}
\tablecaption{\label{tab:ufits}Parameters for skewed-uniform fits.}
\tablehead{
          & \colhead{$\Delta\mathrm{UV}_c$} & \colhead{$w_{\Delta\mathrm{UV}}$} &  \colhead{skew}
}
\startdata
Fornax 1 & $0.028^{+0.027}_{-0.026}$ & $0.213^{+0.051}_{-0.050}$ & \phs$0.454^{+0.395}_{-0.750}$ \\
Fornax 2 & $0.098^{+0.015}_{-0.017}$ & $0.193^{+0.029}_{-0.025}$ & $-0.418^{+0.516}_{-0.405}$ \\
Fornax 3 & $0.060^{+0.018}_{-0.016}$ & $0.172^{+0.029}_{-0.025}$ & \phs$0.344^{+0.449}_{-0.651}$ \\
Fornax 5 & $0.102^{+0.018}_{-0.017}$ & $0.185^{+0.034}_{-0.027}$ & \phs$0.374^{+0.442}_{-0.597}$
\enddata
\tablecomments{
$\Delta$UV$_c$ is the center of the model distribution in $\Delta$(F343N-F555W), $w_{\Delta\mathrm{UV}}$ the width, and ``skew'' the skewness parameter. See Section~\ref{sec:umcmc} for details.
}
\end{deluxetable}

While convenient, the description of the intrinsic color distributions as double-Gaussians is somewhat artificial. As an alternative, we also tried modeling the color distributions as ``skewed uniform'' distributions, described as
\begin{equation}
  \mathcal{U}(\Delta\mathrm{UV}) = \frac{1 + 2 \times \mathrm{skew} \times (\Delta\mathrm{UV} - \Delta\mathrm{UV}_c)/w_{\Delta\mathrm{UV}}}{w_{\Delta\mathrm{UV}}}  
  \label{eq:uskew}
\end{equation}
for  $\Delta\mathrm{UV}_c - w_{\Delta\mathrm{UV}}/2 < \Delta\mathrm{UV} < \Delta\mathrm{UV}_c + w_{\Delta\mathrm{UV}}/2$ and $\mathcal{U}(\Delta\mathrm{UV}) = 0$ otherwise. For skew=0, $\mathcal{U}$ is simply a box function of width $w_{\Delta\mathrm{UV}}$, centered at $\Delta\mathrm{UV}_c$. For skew $\neq0$ the ``top'' of the box is tilted so that for $\mathrm{skew}=1$, the function becomes triangular with $\mathcal{U}(\Delta\mathrm{UV}_c - w_{\Delta\mathrm{UV}}/2)=0$ and 
$\mathcal{U}(\Delta\mathrm{UV}_c + w_{\Delta\mathrm{UV}}/2) = 2/w_{\Delta\mathrm{UV}}$. 
A positive skew parameter thus corresponds to a color distribution weighted towards redder (more N-enhanced) stars.
We allowed the ``skew'' parameter to have values in the range $-1 < \mathrm{skew} < +1$. We then convolved the intrinsic model color distributions described by Eq.~(\ref{eq:uskew}) with the error distributions determined from the artificial star tests and solved for the three parameters ($\Delta\mathrm{UV}_c$, $w_{\Delta\mathrm{UV}}$, and skew) in a manner similar to that described in Section~\ref{sec:mcmc}. The resulting median parameter values and the 16\% and 84\% percentiles are listed in Table~\ref{tab:ufits} and the model color distributions for the median parameter values are shown as (blue) dashed lines in Fig.~\ref{fig:uvhist}.

When convolved with the error distributions, the best-fitting skewed uniform distributions are very similar to the double-Gaussian models. From Table~\ref{tab:ufits}, we see that the preferred value of the skew parameter is negative for Fornax~2 (indicating a larger fraction of N-normal stars) whereas we find positive skew parameters for Fornax~1, 3, and 5. This is consistent with the results of the double-Gaussian fits, in which the bluer component was found to dominate in Fornax~2. 
While the skewness of the color distributions is poorly constrained and all clusters are consistent with flat distributions ($\mathrm{skew}=0$), the widths of the distributions, $w_{\Delta\mathrm{UV}}$, are fairly well constrained and fall in the range 0.17--0.21 mag. This is very similar to our estimates from Section~\ref{sec:cspreads}.

\subsection{Comparison with M15}
\label{sec:m15cmp}

As noted above, the well-studied Galactic GC M15 was included as a comparison target. Unlike the Fornax GCs, the  errors are very small for the M15 photometry ($\sim0.01$ mag), and this must be taken into account when carrying out a detailed comparison. We first defined the $\Delta$(F343N-F555W) index for M15 in the same way as for the Fornax GCs. Measuring the dispersion, we found $\sigma_\mathrm{obs,UV} = 0.052$ mag for M15, very similar to the \emph{intrinsic} spreads estimated for the Fornax GCs  (Sect.~\ref{sec:cspreads}). 

In order to compare more directly with the Fornax data, we convolved the M15 $\Delta$(F343N-F555W) distribution with the error distributions determined from the artificial star tests for each Fornax cluster. The results are shown in Fig.~\ref{fig:uvhist} as black dotted lines. We see that, in general, the error-convolved M15 color distributions bear close resemblance to the best-fitting double-Gaussian or skewed uniform distributions. The standard deviations of the error-convolved M15 color distributions are $\sigma_\mathrm{M15} = 0.065$ mag, 0.079 mag, 0.074 mag and 0.075 mag
when convolved with the error distributions for Fornax~1, Fornax~2, Fornax~3, and Fornax~5, respectively. 
Note that a small correction ($\sim0.01$ mag) should, in principle, be made for the contribution from photometric uncertainties in M15; this should be subtracted from the dispersions quoted above.
Nevertheless, the dispersions are very similar to those listed in Table~\ref{tab:spread} for the Fornax GCs and a Levene test shows that the small differences could easily arise by chance, with $P>0.4$ in all cases. 
 We therefore conclude that \emph{the observed $\Delta$(F343N-F555W) color distributions of the Fornax GCs are fully consistent with the corresponding color distribution measured for M15}, convolved with the observational errors of the Fornax data as determined from the artificial star tests. 

\subsection{Radial distributions}

\begin{figure}
\includegraphics[width=\columnwidth]{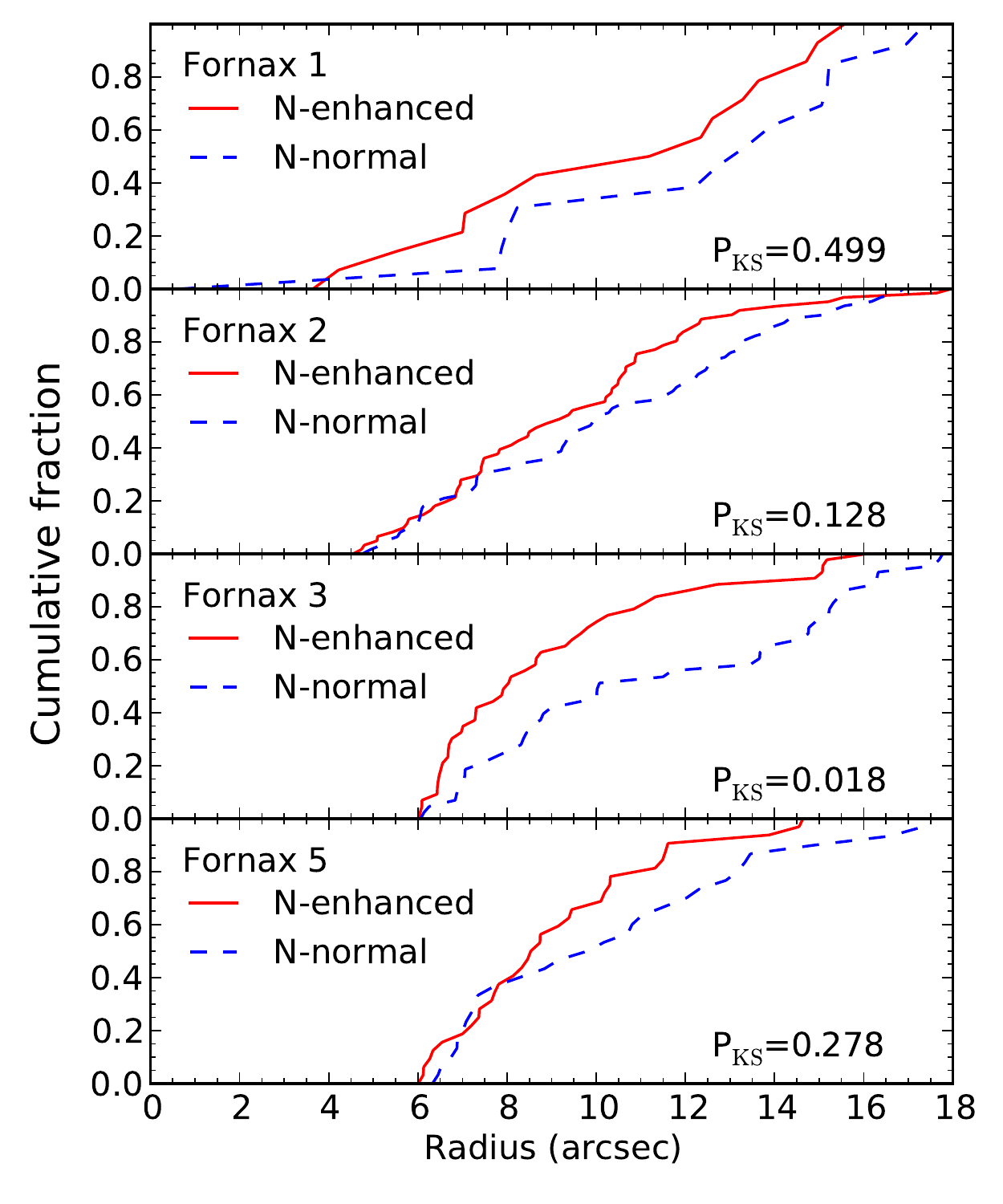}
\caption{\label{fig:rdist}Cumulative radial distributions of N-normal and N-enhanced stars.\\}
\end{figure}

To investigate the radial distributions of stars with different chemical composition, we divided the RGB stars in each cluster into two equal-sized sub-samples based on their $\Delta$(F343N-F555W) colors. The ``first'' generation stars were thus defined as stars with $\Delta$(F343N-F555W) $<$ med($\Delta$F343N-F555W) while the ``second'' generation stars were defined as stars with $\Delta$(F343N-F555W) $\geq$ med($\Delta$F343N-F555W). Figure~\ref{fig:rdist} shows the cumulative radial distributions of the two sub-samples of stars. In each panel we indicate the $P$ value from a two-sample Kolmogorov-Smirnov test, corresponding to the null hypothesis that the two radial distributions are drawn from the same parent distribution.

Since stars in the central regions of the clusters are missing from our samples (except in Fornax~1), it is not possible to make a quantitative comparison of the radial distributions in terms of, say, half-mass radii.  Nevertheless, we see that the second-generation (N-enhanced) stars tend to be more centrally concentrated in all four clusters, although the differences are not highly significant for any individual cluster. 
The inner radii of $4\farcs5$ (Fornax~2) and $6\farcs0$ (Fornax~3 and 5) correspond to linear scales of 3 pc and 4 pc, respectively, at the distance of the Fornax dSph. For comparison, the half-light radii are $18\arcsec$ (Fornax~1), $12\farcs5$ (Fornax~2), $8\farcs2$ (Fornax~3) and $9\farcs6$ (Fornax~5), using the structural parameters from \citet{Mackey2003a}. In terms of the half-light radii, the radial ranges covered here are then quite similar those of \citet{Lardo2010} who studied the radial distributions of stars in Galactic GCs using SDSS photometry. 
These authors also found the reddest RGB stars (using Sloan $u-g$ colors) to be more centrally concentrated, i.e., a similar result to that found here for the Fornax GCs. A tendency for the ``second-generation'' stars to be more centrally concentrated has also been found by other authors \citep{Carretta2009a,Carretta2010b,Kravtsov2010,Milone2012a,Kravtsov2014}, and is also expected in most theoretical scenarios for the origin of multiple populations in GCs \citep{DErcole2008,Bastian2013a,Krause2013,Vesperini2013}.

\section{Discussion}

\subsection{Multiple populations in the Fornax GCs}

The analysis in the previous sections indicates that the $\Delta$(F343N-F555W) color spreads of RGB stars in the Fornax GCs are similar to that in M15 and that
all four metal-poor Fornax GCs contain roughly equal numbers of ``normal'' and N-enhanced stars. This agrees well with the analysis of \citet{DAntona2013}, who estimated that Fornax 2, 3, and 5 contain 54\%-65\% second-generation stars, based on modeling of the horizontal branch morphology.
Since we do not probe the central regions of most of the clusters, and the N-enhanced stars appear to be more centrally concentrated, our estimated first-generation fractions should probably be considered upper limits. This should be kept in mind when comparing the ratios found here with observations of other clusters that may cover different radial ranges. The exact division between N-normal and N-enhanced stars is clearly somewhat arbitrary, since we cannot tell from our observations whether the stars are really divided into two distinct groups. Nevertheless, the conclusion that second-generation stars constitute a significant, and possibly dominant, fraction of the stars in the Fornax GCs, is in agreement with the large amount of work done on Milky Way GCs \citep{Gratton2012}.

The estimated range of about 2 dex in [N/Fe] in the Fornax GCs is also similar to that typical of Galactic GCs. \citet{Yong2008} found star-to-star variations of 1.95 dex in $\mathrm{[N/Fe]}$ in NGC~6752, and listed several other GCs with similar nitrogen abundance spreads. More recently, the SUMO project \citep{Monelli2013} has measured the light element abundance-sensitive $c_{U,B,I}$ index for RGB stars in 23 Galactic GCs. All of these clusters show a spread in the $c_{U,B,I}$ index similar to that seen in NGC~6752, suggesting a similar range of light element abundance variations.

Relatively little is known about abundance variations and multiple stellar populations in other extragalactic GCs.  At least some GCs in the Large Magellanic Cloud display Na-O and Mg-Al anti-correlations similar to those observed in Galactic GCs \citep{Mucciarelli2009}. From their spectroscopy of three stars in each of the clusters Fornax 1, 2, and 3, \citet{Letarte2006} found significant spreads in $\mathrm{[Mg/Fe]}$ and $\mathrm{[Na/Fe]}$ in Fornax~1 and Fornax~3.
Only two of those stars overlap with our dataset (D164 in Fornax~1, with ID 2966 in Table~\ref{tab:f1}, and B226 in Fornax~2 with ID 54687 in Table~\ref{tab:f2}). Both have $M_V<-2$ and are located near the tip of the RGB, where our photometry is less sensitive to N abundance variations and deep mixing is likely to have modified the N abundances. In any case, \citet{Letarte2006} did not measure N for these stars and a direct comparison with our photometry is, therefore, not possible.
 From integrated-light spectroscopy of Fornax 3, 4, and 5, we found that the $\mathrm{[Mg/Fe]}$ ratios were lower than the  $\mathrm{[Ca/Fe]}$ and  $\mathrm{[Ti/Fe]}$ ratios, possibly an indication that the Mg-Al anticorrelation is present in these clusters \citep{Larsen2012a}. Similar results have been found from integrated-light spectroscopy of GCs in M31 \citep{Colucci2009} and in the WLM galaxy \citep{Larsen2014}. In the WLM GC we also found an enhanced [Na/Fe] ratio, again consistent with the patterns observed in Galactic GCs. 
However, the interpretation of the integrated-light measurements is not straight forward. In particular, it is unclear why depleted $\mathrm{[Mg/Fe]}$ ratios tend to be seen more frequently in integrated-light measurements than in observations of individual stars in GCs \citep{Larsen2014}.

The cluster Fornax~1 remains a puzzle. The very low metallicity and old age of this cluster are difficult to reconcile with the red horizontal branch morphology, especially if a (He-enriched) second generation is present in the cluster. (Note that, even if we omit the selection on rms$_\mathrm{F343N}$, the CMD contains no additional HB stars bluewards of those seen in Fig.~\ref{fig:vif}.).
In the  \citet{DAntona2008} model for HB morphology, the extended blue horizontal branches of GCs arise from the faster evolution of He-enriched second-generation stars, which are expected to have lower envelope masses on the HB compared to first-generation stars. 
 \citet{DAntona2013} thus suggested that the red HB morphology of Fornax~1 might be explained if it is a ``first-generation only'' cluster with a slightly higher metallicity than that found by \citet{Letarte2006}. 
However, the evidence for light-element abundance spreads presented here  \citep[and the detailed abundance measurements of][]{Letarte2006} suggests that Fornax~1 hosts the usual proportion of chemically anomalous stars, although there are currently no direct measurements of the He abundance of Fornax~1 (or indeed any of the other GCs in Fornax). While there is observational support for a correlation between location on the HB and He abundance in some Milky Way GCs \citep{DAlessandro2013}, surface He abundances may be modified by stellar evolutionary effects, and  measuring He for stars of different effective temperatures along the HB is difficult \citep{Valcarce2014}. Alternatively, a redder HB might result if Fornax 1 is significantly younger than the other Fornax GCs, but this seems to be ruled out by the main sequence turn-off location which is similar to that of the other clusters \citep[Fig.~\ref{fig:vif}; see also][]{Buonanno1998}.

\subsection{Metallicities}

\begin{figure}
\includegraphics[width=\columnwidth]{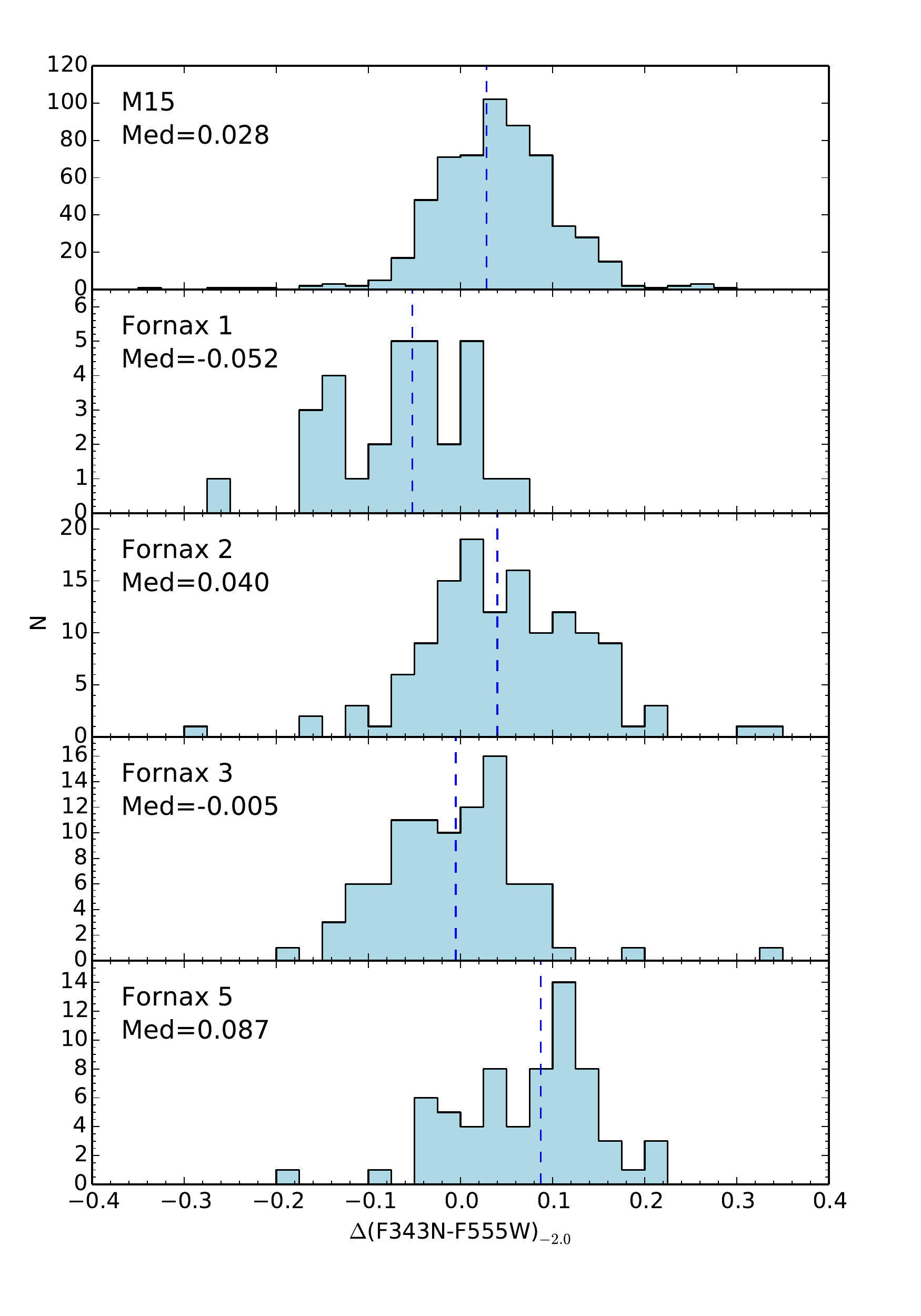}
\caption{\label{fig:uvh_ifix}Comparison of $\Delta$(F343N-F555W) distributions relative to an isochrone of a fixed metallicity of $\mathrm{[Fe/H]}=-2.0$ and N-normal composition.}
\end{figure}

We finally revisit the question of the metallicities of the Fornax clusters. In addition to being sensitive to [N/Fe], the F343N-F555W colors are also sensitive to the overall metallicity and vary by 0.09 mag between  $\mathrm{[Fe/H]}=-2.5$ and $\mathrm{[Fe/H]}=-2.0$ for RGB stars at $M_V\approx+2$. The corresponding variation in F555W-F814W is only 0.01 mag. The use of F343N-F555W as a metallicity indicator is complicated by the additional sensitivity to the light elements, but we may still gain some insight into  metallicity variations from the overall shifts of the color distributions.
Figure~\ref{fig:uvh_ifix} again shows a comparison of the $\Delta$(F343N-F555W) distributions in the clusters, but now defined with respect to an isochrone of a fixed metallicity of $\mathrm{[Fe/H]}=-2.0$. We here use model colors calculated for a N-normal composition. 
We also include the M15 data in this figure, noting that the F555W model colors for M15 were computed specifically for the WFC3/F555W filter, whose transmission curve differs somewhat from that of the WFPC2/F555W filter used for the Fornax observations.

The comparison in Fig.~\ref{fig:uvh_ifix} supports the previous findings that Fornax~1 is the most metal-poor of the Fornax GCs, followed by Fornax~3, Fornax~2, and Fornax~5 (in that order). Fornax~4, which is not included here, is by far the most metal-rich of the clusters. While we have used isochrones with $[\alpha/\mathrm{Fe}] = +0.2$ for the figure, $[\alpha/\mathrm{Fe}] = +0.4$ may be more appropriate for M15 \citep{Sneden1997,Roediger2014}. The model colors then shift to the red by about 0.02 mag and the median color offset of the M15 RGB stars would then be 0.011 mag. In any case, the metallicity of M15 \citep[$\mathrm{[Fe/H]}=-2.3$;][]{Carretta2009c} appears to be intermediate between those of Fornax~2 and Fornax~3. 
We have assumed $A_V=0.300$ mag for M15, which is about 0.06 mag less than the older value of \citet{Schlegel1998} tabulated in NED. This corresponds to a difference of 0.034 mag in F343N-F555W. 
If the true extinction towards M15 is slightly higher than the value we have assumed, then the median F343N-F555W colors of the M15 RGB stars would become very similar to those in Fornax~3 but would remain redder than those in Fornax~1, even when accounting for small zero-point uncertainties in the F343N-F555W colors (Sect.~\ref{sec:cspreads}).  The reddening towards the Fornax GCs is very small, but nevertheless also subject to some uncertainty. Other literature values tend to be higher than those assumed here, and would shift the Fornax GCs further to the left in Fig.~\ref{fig:uvh_ifix}. For example, \citet{Buonanno1998} find $E(V\!-\!I)$ in the range 0.05 - 0.09 mag, about twice as large as the \citet{Schlafly2011} values. In any case, this comparison shows that the Fornax GCs are indeed very metal-poor, with Fornax~1 and Fornax~3 having metallicities similar to, or below, that of M15, in agreement with the high-dispersion spectroscopy \citep{Letarte2006,Larsen2012a}.

Concerning the field stars, the comparison of field star and GC metallicities in \citet{Larsen2012} was based on the Ca {\sc ii} IR triplet metallicity measurements for field stars of \citet{Battaglia2006}. \citet{DAntona2013} expressed the concern that the field star and GC metallicities might not be on the same scale. The Ca {\sc ii} triplet scale was indeed revised by \citet{Starkenburg2010}, but this mainly affects stars with $\mathrm{[Fe/H]}<-2$. In particular, the metallicity distribution of the Fornax dSph changes very little when applying the more recent calibration of \citet[][their Fig.~13]{Starkenburg2010}.
Furthermore, \citet{Starkenburg2010} find that the metallicities of stars measured with their Ca {\sc ii} triplet calibration agree very well with measurements of [Fe/H] from high-dispersion spectroscopy over a wide metallicity range. Recently, \citet{Hendricks2014} have applied the Ca {\sc ii} triplet technique to individual stars in Fornax~2 and Fornax~5 and report $\mathrm{[Fe/H]}=-2.04\pm0.04$ and $\mathrm{[Fe/H]}=-2.02\pm0.11$ for the two clusters, respectively. These values are, again, in excellent agreement with those derived from high-dispersion spectroscopy.
The metallicities of GCs and field stars in Fornax therefore appear to be consistent with a single scale that agrees with direct measurements of [Fe/H] from high-dispersion spectroscopy.

In conclusion, then, the Fornax GCs are very similar to Milky Way GCs in terms of their stellar population properties. 
Consequently, scenarios that aim to explain the presence of chemical abundance anomalies must apply equally well to clusters in these different environments. The dwarf galaxies provide a particularly stringent constraint on scenarios that require a large amount of mass loss, due to the relatively high fractions of metal-poor stars that belong to clusters in these galaxies \citep{Larsen2012,Larsen2014}.

\section{Conclusions}

We have presented new observations of the four most metal-poor globular clusters in the Fornax dwarf galaxy, obtained with the F343N filter on the Wide Field Camera 3. By combining these observations with archival data in F555W and F814W, we have looked for variations in the nitrogen abundances of red giants in the clusters. Our main findings are as follows:

\begin{itemize}
\item The observed colors of stars on the lower RGB are consistent with the overall low metallicities ($\mathrm{[Fe/H]}<-2$) previously determined from high-dispersion spectroscopy. Fornax~1 and 3 are the most metal-poor of the clusters with $\mathrm{[Fe/H]}$ at least as low as that of M15, while Fornax~2 and 5 have slightly higher metallicities.
The F343N-F555W color distributions are consistent with the claim \citep{Letarte2006} that Fornax~1 ``holds the record for the lowest metallicity globular cluster''.
\item All four GCs display a spread in the F343N-F555W colors of RGB stars that is consistent with a range in N abundances of about 2 dex. This is similar to the spread seen in Milky Way globular clusters. The F555W-F814W colors, instead, show no spread beyond the measurement errors.
\item We model the observed color distributions as double Gaussians convolved with the error distributions as determined from artificial star tests. The color distributions of all four clusters can be described as a sum of two Gaussian components with roughly equal weights, suggesting roughly similar numbers of first- (N-normal) and second-generation (N-enhanced) stars. Formally, we find the N-normal fractions to be $0.39^{+0.31}_{-0.24}$ in Fornax~1, $0.65^{+0.14}_{-0.16}$ in Fornax~2, $0.43^{+0.32}_{-0.21}$ in Fornax~3, and $0.37^{+0.15}_{-0.13}$ in Fornax~5.
\item The observed spreads in the F343N-F555W colors of the Fornax GCs are consistent with the corresponding spread for M15, after convolving the latter with the error distributions of the Fornax clusters.
\item The radial distributions of the N-normal stars appear to be more extended than those of the N-enhanced stars, although this result is only of marginal statistical significance. Since we do not probe the central regions of the clusters (except Fornax 1), the global N-normal fractions are therefore likely to be lower than the numbers quoted above.
\end{itemize}

We conclude that the Fornax GCs are similar to Galactic GCs in terms of their stellar population properties. At least half  of the stars in the clusters appear to have formed from material that was enriched by proton-capture nucleosynthesis. The same processes that were responsible for the chemical anomalies observed in Galactic GCs are thus likely to have operated in the Fornax clusters. 
The implication is that theoretical scenarios for the origin of multiple populations in GCs must account not only for the usual mass budget problem (i.e., the large fractions of polluted stars in GCs) but also for the ``external mass budget'' problem \citep{Bastian2013a} that arises from the high ratio of metal-poor GCs vs.\ field stars in dwarf galaxies. 
Since 1/5-1/4 of the metal-poor field stars in the Fornax dSph belong to the four metal-poor GCs, with similar or even more extreme ratios in the WLM and IKN dwarfs \citep{Larsen2012,Larsen2014}, there is a clear tension between these observations and scenarios for GC formation that require GCs to have lost a factor of 10 or more of their original mass. 
More generally, this also constrains the fraction of metal-poor stars that could have formed in the field or in disrupted clusters.

\acknowledgements
JB and JS acknowledge support for HST Program number GO-13295 from NASA through grants HST-GO-13295.02 and HST-GO-13295.03 from the Space Telescope Science Institute, which is operated by the Association of Universities for Research in Astronomy, Incorporated, under NASA contract NAS5-26555. JB also acknowledges HST grant HST-GO-13048.02 and NSF grant AST-1109878.
Funding for the Stellar Astrophysics Centre is provided by The Danish National Research Foundation. The research is supported by the ASTERISK project (ASTERoseismic Investigations with SONG and Kepler) funded by the European Research Council (Grant agreement no.: 267864).
We thank Ricardo Salinas for help with the data analysis,  Aaron Dotter for discussion about the horizontal branch morphology of Fornax~1 and Antonino Milone for discussion about photometry. The anonymous referee is thanked for a prompt and helpful report.
This research has made use of the NASA/IPAC Extragalactic Database (NED), which is operated by the Jet Propulsion Laboratory, California Institute of Technology, under contract with the National Aeronautics and Space Administration.

Facilities: \facility{HST(WFC3)}, \facility{HST(WFPC2)}

\bibliographystyle{apj}
\bibliography{libmen.bib}

\clearpage

\LongTables

\begin{deluxetable*}{lcccrrrrrrrrrr}
\tablecaption{\label{tab:modcol}Model colors for RGB stars.}
\tablecolumns{15}
\tabletypesize{\scriptsize}
\tablehead{
\colhead{[Fe/H]} & \colhead{[$\alpha$/Fe]} & \colhead{$T_\mathrm{eff}$} & \colhead{$\log g$} &
 \multicolumn{5}{c}{N-normal} & \multicolumn{5}{c}{N-enhanced} \\
 & & & & 343$_3$ & 555$_3$ & 814$_3$ & 555$_2$ & 814$_2$ &
343$_3$ & 555$_3$ & 814$_3$ & 555$_2$ & 814$_2$
}
\startdata
$-2.0$ & $+0.2$ & 6602 & 4.19 & 4.015 & 3.896 & 4.675 & 3.933 & 4.667 & 4.044 & 3.895 & 4.674 & 3.932 & 4.666 \\
$-2.0$ & $+0.2$ & 6598 & 4.12 & 3.834 & 3.694 & 4.474 & 3.731 & 4.467 & 3.861 & 3.692 & 4.473 & 3.730 & 4.466 \\
$-2.0$ & $+0.2$ & 6525 & 4.02 & 3.654 & 3.496 & 4.262 & 3.532 & 4.254 & 3.685 & 3.495 & 4.261 & 3.531 & 4.253 \\
$-2.0$ & $+0.2$ & 6325 & 3.89 & 3.476 & 3.304 & 4.021 & 3.337 & 4.014 & 3.527 & 3.302 & 4.020 & 3.335 & 4.013 \\
$-2.0$ & $+0.2$ & 5882 & 3.67 & 3.361 & 3.122 & 3.724 & 3.146 & 3.717 & 3.478 & 3.118 & 3.721 & 3.142 & 3.714 \\
$-2.0$ & $+0.2$ & 5609 & 3.50 & 3.266 & 2.929 & 3.455 & 2.947 & 3.449 & 3.422 & 2.925 & 3.453 & 2.944 & 3.446 \\
$-2.0$ & $+0.2$ & 5512 & 3.39 & 3.117 & 2.731 & 3.229 & 2.747 & 3.223 & 3.285 & 2.728 & 3.227 & 2.744 & 3.221 \\
$-2.0$ & $+0.2$ & 5440 & 3.24 & 2.871 & 2.432 & 2.909 & 2.447 & 2.903 & 3.041 & 2.429 & 2.907 & 2.443 & 2.901 \\
$-2.0$ & $+0.2$ & 5365 & 3.01 & 2.433 & 1.933 & 2.387 & 1.946 & 2.381 & 2.604 & 1.930 & 2.385 & 1.943 & 2.380 \\
$-2.0$ & $+0.2$ & 5288 & 2.78 & 2.003 & 1.435 & 1.864 & 1.445 & 1.858 & 2.175 & 1.432 & 1.862 & 1.443 & 1.857 \\
$-2.0$ & $+0.2$ & 5205 & 2.55 & 1.585 & 0.938 & 1.338 & 0.946 & 1.332 & 1.757 & 0.935 & 1.337 & 0.943 & 1.332 \\
$-2.0$ & $+0.2$ & 5112 & 2.31 & 1.206 & 0.441 & 0.805 & 0.446 & 0.800 & 1.375 & 0.438 & 0.805 & 0.443 & 0.800 \\
$-2.0$ & $+0.2$ & 5012 & 2.07 & 0.834 & -0.056 & 0.267 & -0.054 & 0.262 & 0.999 & -0.058 & 0.268 & -0.056 & 0.263 \\
$-2.0$ & $+0.2$ & 4908 & 1.82 & 0.516 & -0.550 & -0.278 & -0.554 & -0.283 & 0.672 & -0.553 & -0.277 & -0.556 & -0.282 \\
$-2.0$ & $+0.2$ & 4790 & 1.56 & 0.220 & -1.045 & -0.833 & -1.054 & -0.838 & 0.363 & -1.049 & -0.830 & -1.057 & -0.835 \\
$-2.0$ & $+0.4$ & 6547 & 4.18 & 4.023 & 3.907 & 4.672 & 3.943 & 4.664 & 4.057 & 3.906 & 4.671 & 3.942 & 4.663 \\
$-2.0$ & $+0.4$ & 6517 & 4.10 & 3.842 & 3.707 & 4.467 & 3.742 & 4.459 & 3.876 & 3.705 & 4.466 & 3.741 & 4.458 \\
$-2.0$ & $+0.4$ & 6397 & 3.98 & 3.662 & 3.510 & 4.243 & 3.544 & 4.235 & 3.707 & 3.508 & 4.241 & 3.542 & 4.234 \\
$-2.0$ & $+0.4$ & 6106 & 3.82 & 3.513 & 3.317 & 3.978 & 3.345 & 3.971 & 3.594 & 3.314 & 3.976 & 3.343 & 3.969 \\
$-2.0$ & $+0.4$ & 5676 & 3.60 & 3.434 & 3.132 & 3.676 & 3.151 & 3.670 & 3.584 & 3.128 & 3.674 & 3.147 & 3.667 \\
$-2.0$ & $+0.4$ & 5536 & 3.47 & 3.300 & 2.934 & 3.438 & 2.950 & 3.432 & 3.468 & 2.930 & 3.436 & 2.947 & 3.429 \\
$-2.0$ & $+0.4$ & 5468 & 3.37 & 3.143 & 2.736 & 3.220 & 2.751 & 3.214 & 3.317 & 2.733 & 3.218 & 2.748 & 3.212 \\
$-2.0$ & $+0.4$ & 5408 & 3.22 & 2.891 & 2.436 & 2.903 & 2.450 & 2.897 & 3.065 & 2.433 & 2.902 & 2.447 & 2.896 \\
$-2.0$ & $+0.4$ & 5335 & 3.00 & 2.453 & 1.937 & 2.380 & 1.948 & 2.375 & 2.627 & 1.934 & 2.379 & 1.946 & 2.373 \\
$-2.0$ & $+0.4$ & 5254 & 2.76 & 2.027 & 1.439 & 1.856 & 1.449 & 1.850 & 2.202 & 1.436 & 1.855 & 1.446 & 1.849 \\
$-2.0$ & $+0.4$ & 5171 & 2.53 & 1.626 & 0.939 & 1.326 & 0.946 & 1.321 & 1.798 & 0.936 & 1.326 & 0.944 & 1.320 \\
$-2.0$ & $+0.4$ & 5072 & 2.29 & 1.240 & 0.443 & 0.792 & 0.447 & 0.787 & 1.410 & 0.441 & 0.792 & 0.445 & 0.787 \\
$-2.0$ & $+0.4$ & 4968 & 2.04 & 0.900 & -0.054 & 0.249 & -0.054 & 0.244 & 1.063 & -0.056 & 0.250 & -0.056 & 0.245 \\
$-2.0$ & $+0.4$ & 4856 & 1.78 & 0.574 & -0.549 & -0.300 & -0.554 & -0.305 & 0.727 & -0.553 & -0.298 & -0.557 & -0.302 \\
$-2.0$ & $+0.4$ & 4731 & 1.52 & 0.335 & -1.041 & -0.863 & -1.053 & -0.867 & 0.469 & -1.046 & -0.859 & -1.058 & -0.864 \\
$-2.1$ & $+0.2$ & 6620 & 4.20 & 4.011 & 3.898 & 4.680 & 3.935 & 4.672 & 4.034 & 3.897 & 4.679 & 3.934 & 4.672 \\
$-2.1$ & $+0.2$ & 6622 & 4.12 & 3.830 & 3.694 & 4.480 & 3.732 & 4.473 & 3.852 & 3.693 & 4.480 & 3.731 & 4.472 \\
$-2.1$ & $+0.2$ & 6555 & 4.03 & 3.650 & 3.497 & 4.269 & 3.533 & 4.261 & 3.674 & 3.496 & 4.268 & 3.532 & 4.261 \\
$-2.1$ & $+0.2$ & 6368 & 3.90 & 3.470 & 3.305 & 4.033 & 3.338 & 4.025 & 3.510 & 3.304 & 4.031 & 3.337 & 4.024 \\
$-2.1$ & $+0.2$ & 5951 & 3.69 & 3.339 & 3.120 & 3.741 & 3.145 & 3.734 & 3.434 & 3.117 & 3.738 & 3.142 & 3.731 \\
$-2.1$ & $+0.2$ & 5633 & 3.51 & 3.244 & 2.929 & 3.462 & 2.948 & 3.456 & 3.390 & 2.926 & 3.460 & 2.944 & 3.453 \\
$-2.1$ & $+0.2$ & 5526 & 3.39 & 3.098 & 2.733 & 3.234 & 2.749 & 3.228 & 3.258 & 2.729 & 3.232 & 2.746 & 3.226 \\
$-2.1$ & $+0.2$ & 5449 & 3.24 & 2.853 & 2.434 & 2.913 & 2.448 & 2.907 & 3.017 & 2.430 & 2.911 & 2.445 & 2.905 \\
$-2.1$ & $+0.2$ & 5372 & 3.01 & 2.414 & 1.935 & 2.390 & 1.948 & 2.385 & 2.581 & 1.932 & 2.389 & 1.945 & 2.383 \\
$-2.1$ & $+0.2$ & 5294 & 2.78 & 1.983 & 1.436 & 1.867 & 1.447 & 1.861 & 2.152 & 1.433 & 1.866 & 1.444 & 1.860 \\
$-2.1$ & $+0.2$ & 5213 & 2.55 & 1.562 & 0.938 & 1.340 & 0.946 & 1.335 & 1.731 & 0.935 & 1.339 & 0.944 & 1.334 \\
$-2.1$ & $+0.2$ & 5120 & 2.31 & 1.181 & 0.440 & 0.808 & 0.446 & 0.802 & 1.348 & 0.438 & 0.807 & 0.444 & 0.802 \\
$-2.1$ & $+0.2$ & 5022 & 2.07 & 0.805 & -0.056 & 0.270 & -0.054 & 0.265 & 0.970 & -0.058 & 0.270 & -0.056 & 0.265 \\
$-2.1$ & $+0.2$ & 4918 & 1.82 & 0.485 & -0.550 & -0.274 & -0.553 & -0.279 & 0.643 & -0.552 & -0.273 & -0.555 & -0.278 \\
$-2.1$ & $+0.2$ & 4803 & 1.56 & 0.184 & -1.045 & -0.829 & -1.053 & -0.834 & 0.331 & -1.048 & -0.827 & -1.056 & -0.832 \\
$-2.2$ & $+0.4$ & 6599 & 4.19 & 4.012 & 3.906 & 4.682 & 3.943 & 4.675 & 4.033 & 3.905 & 4.682 & 3.942 & 4.674 \\
$-2.2$ & $+0.4$ & 6584 & 4.11 & 3.832 & 3.705 & 4.481 & 3.742 & 4.473 & 3.853 & 3.704 & 4.480 & 3.741 & 4.473 \\
$-2.2$ & $+0.4$ & 6495 & 4.01 & 3.650 & 3.507 & 4.263 & 3.542 & 4.256 & 3.675 & 3.505 & 4.262 & 3.541 & 4.255 \\
$-2.2$ & $+0.4$ & 6263 & 3.86 & 3.472 & 3.315 & 4.015 & 3.347 & 4.008 & 3.518 & 3.314 & 4.014 & 3.345 & 4.006 \\
$-2.2$ & $+0.4$ & 5807 & 3.64 & 3.374 & 3.127 & 3.709 & 3.150 & 3.702 & 3.484 & 3.124 & 3.706 & 3.147 & 3.700 \\
$-2.2$ & $+0.4$ & 5591 & 3.49 & 3.252 & 2.935 & 3.454 & 2.952 & 3.448 & 3.399 & 2.931 & 3.452 & 2.949 & 3.445 \\
$-2.2$ & $+0.4$ & 5505 & 3.38 & 3.097 & 2.736 & 3.230 & 2.752 & 3.224 & 3.254 & 2.732 & 3.228 & 2.748 & 3.222 \\
$-2.2$ & $+0.4$ & 5435 & 3.23 & 2.847 & 2.436 & 2.911 & 2.450 & 2.905 & 3.008 & 2.433 & 2.909 & 2.447 & 2.903 \\
$-2.2$ & $+0.4$ & 5359 & 3.00 & 2.408 & 1.937 & 2.388 & 1.949 & 2.382 & 2.571 & 1.934 & 2.386 & 1.946 & 2.380 \\
$-2.2$ & $+0.4$ & 5279 & 2.77 & 1.978 & 1.439 & 1.864 & 1.449 & 1.858 & 2.144 & 1.436 & 1.862 & 1.446 & 1.856 \\
$-2.2$ & $+0.4$ & 5196 & 2.54 & 1.573 & 0.938 & 1.334 & 0.946 & 1.329 & 1.738 & 0.936 & 1.333 & 0.944 & 1.328 \\
$-2.2$ & $+0.4$ & 5101 & 2.30 & 1.179 & 0.442 & 0.802 & 0.447 & 0.796 & 1.346 & 0.440 & 0.801 & 0.445 & 0.796 \\
$-2.2$ & $+0.4$ & 5000 & 2.05 & 0.806 & -0.055 & 0.261 & -0.054 & 0.256 & 0.972 & -0.057 & 0.261 & -0.056 & 0.256 \\
$-2.2$ & $+0.4$ & 4892 & 1.80 & 0.493 & -0.550 & -0.287 & -0.554 & -0.291 & 0.652 & -0.552 & -0.286 & -0.555 & -0.290 \\
$-2.2$ & $+0.4$ & 4773 & 1.54 & 0.239 & -1.042 & -0.847 & -1.053 & -0.852 & 0.385 & -1.045 & -0.845 & -1.055 & -0.850 \\
$-2.3$ & $+0.2$ & 6652 & 4.20 & 4.005 & 3.900 & 4.689 & 3.938 & 4.682 & 4.020 & 3.900 & 4.689 & 3.937 & 4.681 \\
$-2.3$ & $+0.2$ & 6667 & 4.13 & 3.825 & 3.695 & 4.492 & 3.734 & 4.484 & 3.838 & 3.695 & 4.491 & 3.733 & 4.484 \\
$-2.3$ & $+0.2$ & 6614 & 4.04 & 3.644 & 3.496 & 4.282 & 3.534 & 4.275 & 3.658 & 3.495 & 4.282 & 3.533 & 4.274 \\
$-2.3$ & $+0.2$ & 6450 & 3.92 & 3.461 & 3.304 & 4.052 & 3.339 & 4.044 & 3.484 & 3.303 & 4.051 & 3.338 & 4.043 \\
$-2.3$ & $+0.2$ & 6096 & 3.73 & 3.306 & 3.115 & 3.775 & 3.144 & 3.768 & 3.361 & 3.113 & 3.773 & 3.142 & 3.766 \\
$-2.3$ & $+0.2$ & 5691 & 3.52 & 3.206 & 2.930 & 3.479 & 2.950 & 3.473 & 3.326 & 2.927 & 3.476 & 2.947 & 3.470 \\
$-2.3$ & $+0.2$ & 5558 & 3.40 & 3.061 & 2.733 & 3.243 & 2.750 & 3.237 & 3.201 & 2.729 & 3.241 & 2.747 & 3.235 \\
$-2.3$ & $+0.2$ & 5470 & 3.24 & 2.818 & 2.434 & 2.919 & 2.449 & 2.913 & 2.965 & 2.430 & 2.917 & 2.446 & 2.911 \\
$-2.3$ & $+0.2$ & 5389 & 3.01 & 2.378 & 1.935 & 2.395 & 1.948 & 2.390 & 2.531 & 1.931 & 2.393 & 1.945 & 2.388 \\
$-2.3$ & $+0.2$ & 5311 & 2.79 & 1.943 & 1.436 & 1.871 & 1.447 & 1.866 & 2.101 & 1.433 & 1.870 & 1.444 & 1.864 \\
$-2.3$ & $+0.2$ & 5229 & 2.55 & 1.523 & 0.939 & 1.346 & 0.948 & 1.341 & 1.683 & 0.936 & 1.344 & 0.945 & 1.339 \\
$-2.3$ & $+0.2$ & 5138 & 2.31 & 1.133 & 0.440 & 0.813 & 0.446 & 0.808 & 1.295 & 0.437 & 0.812 & 0.444 & 0.806 \\
$-2.3$ & $+0.2$ & 5041 & 2.07 & 0.752 & -0.056 & 0.277 & -0.054 & 0.271 & 0.915 & -0.058 & 0.276 & -0.056 & 0.271 \\
$-2.3$ & $+0.2$ & 4939 & 1.82 & 0.425 & -0.551 & -0.267 & -0.553 & -0.272 & 0.585 & -0.552 & -0.267 & -0.554 & -0.272 \\
$-2.3$ & $+0.2$ & 4830 & 1.57 & 0.109 & -1.047 & -0.820 & -1.054 & -0.824 & 0.263 & -1.048 & -0.819 & -1.055 & -0.823 \\
$-2.5$ & $+0.2$ & 6676 & 4.21 & 4.000 & 3.901 & 4.695 & 3.939 & 4.688 & 4.009 & 3.901 & 4.695 & 3.939 & 4.687 \\
$-2.5$ & $+0.2$ & 6709 & 4.14 & 3.822 & 3.696 & 4.502 & 3.735 & 4.494 & 3.830 & 3.696 & 4.502 & 3.735 & 4.494 \\
$-2.5$ & $+0.2$ & 6668 & 4.05 & 3.623 & 3.502 & 4.297 & 3.540 & 4.290 & 3.632 & 3.501 & 4.297 & 3.540 & 4.290 \\
$-2.5$ & $+0.2$ & 6528 & 3.94 & 3.456 & 3.302 & 4.069 & 3.338 & 4.061 & 3.468 & 3.301 & 4.068 & 3.338 & 4.061 \\
$-2.5$ & $+0.2$ & 6220 & 3.77 & 3.272 & 3.114 & 3.803 & 3.144 & 3.796 & 3.303 & 3.113 & 3.802 & 3.143 & 3.795 \\
$-2.5$ & $+0.2$ & 5763 & 3.54 & 3.171 & 2.928 & 3.497 & 2.950 & 3.491 & 3.259 & 2.925 & 3.495 & 2.947 & 3.489 \\
$-2.5$ & $+0.2$ & 5592 & 3.40 & 3.029 & 2.732 & 3.253 & 2.750 & 3.246 & 3.146 & 2.729 & 3.250 & 2.747 & 3.244 \\
$-2.5$ & $+0.2$ & 5492 & 3.25 & 2.787 & 2.433 & 2.925 & 2.449 & 2.919 & 2.915 & 2.430 & 2.922 & 2.446 & 2.916 \\
$-2.5$ & $+0.2$ & 5405 & 3.02 & 2.348 & 1.935 & 2.400 & 1.949 & 2.394 & 2.484 & 1.932 & 2.398 & 1.946 & 2.392 \\
$-2.5$ & $+0.2$ & 5327 & 2.79 & 1.911 & 1.436 & 1.876 & 1.448 & 1.871 & 2.053 & 1.433 & 1.874 & 1.445 & 1.869 \\
$-2.5$ & $+0.2$ & 5245 & 2.55 & 1.486 & 0.938 & 1.350 & 0.948 & 1.345 & 1.634 & 0.935 & 1.349 & 0.945 & 1.343 \\
$-2.5$ & $+0.2$ & 5156 & 2.32 & 1.093 & 0.439 & 0.818 & 0.446 & 0.813 & 1.243 & 0.436 & 0.817 & 0.443 & 0.811 \\
$-2.5$ & $+0.2$ & 5061 & 2.08 & 0.705 & -0.057 & 0.284 & -0.053 & 0.279 & 0.860 & -0.059 & 0.283 & -0.055 & 0.278 \\
$-2.5$ & $+0.2$ & 4960 & 1.83 & 0.371 & -0.552 & -0.260 & -0.553 & -0.265 & 0.526 & -0.554 & -0.260 & -0.555 & -0.265 \\
$-2.5$ & $+0.2$ & 4860 & 1.58 & 0.038 & -1.048 & -0.808 & -1.054 & -0.813 & 0.192 & -1.049 & -0.808 & -1.055 & -0.813 
\enddata
\tablecomments{All magnitudes are in the STMAG system. Subscripts $_2$ and $_3$ refer to WFPC2 and WFC3, respectively (e.g., 555$_2$ is the F555W STMAG magnitude for WFPC2).}
\end{deluxetable*}

\end{document}